\documentclass[aps,pra,twocolumn,superscriptaddress]{revtex4-1}

\usepackage{graphicx}  % needed for figures
\usepackage{dcolumn}   % needed for some tables
\usepackage{bm}        % for math
\usepackage{amssymb}   % for math
\usepackage{pdfpages}
\usepackage{hyperref}
\usepackage{amsmath}
\usepackage{latexsym}
\usepackage{epsfig}
\usepackage{array}
\usepackage{ulem}
\usepackage{subfigure}
\usepackage{color}

\begin{document}
\newcommand{\bra}[1]{\mbox{\ensuremath{\langle #1 \vert}}}
\newcommand{\ket}[1]{\mbox{\ensuremath{\vert #1 \rangle}}}
\newcommand{\mb}[1]{\mathbf{#1}}
\newcommand{\phipp}{\big|\phi_{\mb{p}}^{(+)}\big>}
\newcommand{\phipav}{\big|\phi_{\mb{p}}^{\p{av}}\big>}
\newcommand{\pp}[1]{\big|\psi_{p}(#1)\big>}
\newcommand{\drdy}[1]{\sqrt{-R'(#1)}}
\newcommand{\Rb}{$^{87}$Rb}
\newcommand{\kf}{$^{40}$K}
\newcommand{\na}{${^{23}}$Na}
\newcommand{\muK}{\:\mu\textrm{K}}
\newcommand{\p}[1]{\textrm{#1}}
\newcommand\T{\rule{0pt}{2.6ex}}
\newcommand\B{\rule[-1.2ex]{0pt}{0pt}}
\newcommand{\reffig}[1]{\mbox{Fig.~\ref{#1}}}
\newcommand{\refeq}[1]{\mbox{Eq.~(\ref{#1})}}
\hyphenation{Fesh-bach}
\newcommand{\nuke}[1]{}
\newcommand{\note}[1]{\textcolor{red}{[\textrm{#1}]}}

\title{Hong-Ou-Mandel atom interferometry in tunnel-coupled optical tweezers} 

%\author
%{A. M. Kaufman$^{1,2}$, B. J. Lester$^{1,2}$, C. M. Reynolds$^{1,2}$, M. L. Wall$^{1,2}$,\\ M. Foss-Feig$^3$, K. R. A. Hazzard$^{1,2}$,
%A. M. Rey$^{1,2}$, C. A. Regal$^{1,2}$ \\
%\\
%\normalsize{$^1$JILA, National Institute of Standards and Technology and University of Colorado,}\\
%\normalsize{$^2$Department of Physics, University of Colorado, Boulder, Colorado 80309, USA}\\
%\normalsize{$^3$ Joint Quantum Institute and the National Institute of Standards}\\ \normalsize{and Technology, Gaithersburg, Maryland, 20899, USA}\\
%\\
%\normalsize{$^\ast$To whom correspondence should be addressed. E-mail:  adam.kaufman@colorado.edu.}
%}

\author{A. M.~Kaufman}
\affiliation{JILA, National Institute of Standards and Technology and University of Colorado}
\affiliation{Department of Physics, University of Colorado, Boulder, Colorado 80309, USA}
\author{B. J.~Lester}
\affiliation{JILA, National Institute of Standards and Technology and University of Colorado}
\affiliation{Department of Physics, University of Colorado, Boulder, Colorado 80309, USA}
\author{C. M. Reynolds}
\affiliation{JILA, National Institute of Standards and Technology and University of Colorado}
\affiliation{Department of Physics, University of Colorado, Boulder, Colorado 80309, USA}
\author{M. L. Wall}
\affiliation{JILA, National Institute of Standards and Technology and University of Colorado}
\affiliation{Department of Physics, University of Colorado, Boulder, Colorado 80309, USA}
\author{M. Foss-Feig}
\affiliation{Joint Quantum Institute and the National Institute of Standards and Technology, Gaithersburg, Maryland, 20899, USA}
\author{K. R. A. Hazzard}
\affiliation{JILA, National Institute of Standards and Technology and University of Colorado}
\affiliation{Department of Physics, University of Colorado, Boulder, Colorado 80309, USA}
\author{A. M. Rey}
\affiliation{JILA, National Institute of Standards and Technology and University of Colorado}
\affiliation{Department of Physics, University of Colorado, Boulder, Colorado 80309, USA}
\author{C. A. Regal}
\affiliation{JILA, National Institute of Standards and Technology and University of Colorado}
\affiliation{Department of Physics, University of Colorado, Boulder, Colorado 80309, USA}

\date{\today}

\begin{abstract}
The quantum statistics of atoms is typically observed in the behavior of an ensemble via macroscopic observables. However, quantum statistics modifies the behavior of even two particles, inducing remarkable consequences that are at the heart of quantum science. Here we demonstrate near-complete control over all the internal and external degrees of freedom of two laser-cooled $^{87}$Rb atoms trapped in two optical tweezers. This full controllability allows us to implement a massive-particle analog of a Hong-Ou-Mandel interferometer where atom tunneling plays the role of a photon beamsplitter. We use the interferometer to probe the effect of quantum statistics on the two-atom dynamics under tunable initial conditions, chosen to adjust the degree of atomic indistinguishability. Our work thereby establishes laser-cooled atoms in optical tweezers as a new route to bottom-up engineering of scalable, low-entropy quantum systems. 
\end{abstract}
\maketitle 

Quantum interference between possible detection paths for two indistinguishable particles yields information about quantum statistics and correlations~\cite{Glauber1963,Twiss}.  An example is the Hong-Ou-Mandel (HOM) effect, which reveals bosonic quantum statistics through a coalescence effect that causes two indistinguishable photons incident on different ports of a beamsplitter to emerge on the same, yet random, output port~\cite{HOM}.  The HOM effect has been observed with photons~\cite{HOM, Beugnon2006, Walraff2013} and in an analogous experiment with electrons~\cite{electron}, but has never been observed with independently prepared massive bosons.  

Here we demonstrate HOM interferometry with bosonic atoms in tunnel-coupled optical tweezers. We attain the requisite, precise control of the single-atom quantum state by laser cooling each atom to its motional ground state in separated, dynamically positionable optical tweezers. The realization of a low-entropy bosonic state by individually placing atoms in their motional ground state has long been a goal in atomic physics~\cite{Vuletic1998, Weiss2004, MeschedePNAS}. The HOM interference we observe represents a direct observation of quantum indistinguishability with independently prepared, laser-cooled atoms. While the role of quantum statistics in macroscopic ensembles of fermionic and bosonic atoms can be observed via Hanbury Brown and Twiss interference experiments~\cite{Yasuda1996,Folling2005,Schellekens2005,Ottl2005,Rom2006,Jeltes2007,Hodgman2011}, HOM interferometry allows us to study nonclassical few-atom states with single-atom control. Our results lay a foundation for linear quantum computing with atoms~\cite{MilburnRM}, interferometric highly sensitive force detection~\cite{Giovannetti1}, control of neutral atoms in nanoscale optical devices~\cite{Vetsch2010,Thompson2013}, and quantum simulation with laser-cooled atoms in scalable optical tweezer arrays. 

In our work, a double-well trapping potential created with optical tweezers realizes a beamsplitter for single $^{87}\mathrm{Rb}$ atoms.  Analogous to photons incident on separate ports of an HOM beamsplitter, when the two bosonic atoms start in separate wells, a tunnel-coupling can result in the transformation $\vert 1,1 \rangle \rightarrow \frac{1}{\sqrt{2}}(\vert 2, 0\rangle~+~\vert 0,2\rangle)$, where $\vert n_L, n_R \rangle$ is the bosonic state of $n_L$ ($n_R$) atoms in the left (right) tweezer.  To understand the HOM effect in our system, it is helpful to utilize the single-particle states $\ket{L}$ and $\ket{R}$ that correspond to an atom localized in the left or right well.  In this notation, the bose-symmetric state that the atoms initially occupy is $\ket{S} \equiv \frac{1}{\sqrt{2}} ( \ket{L}_1\ket{R}_2+\ket{R}_1\ket{L}_2)$, where the ket subscript is a particle label.  Introducing a tunnel-coupling between the left and right wells allows for the single-atom transformations $\ket{L} \rightarrow \frac{1}{\sqrt{2}}(\ket{L} + i\ket{R})$ and $\ket{R} \rightarrow~\frac{1}{\sqrt{2}}(\ket{R}~+~i\ket{L})$.  One might expect that these transformations yield an equal likelihood of finding the atoms in separate wells or the same well.  However, when the two atoms are indistinguishable, the paths resulting in a single atom in each well destructively interfere (Fig.~1A), and hence one finds $\ket{S} \rightarrow \frac{i}{\sqrt{2}} (\ket{L}_1\ket{L}_2+\ket{R}_1\ket{R}_2 )$.  The atoms, therefore, only appear in the same, yet random, tweezer. 

\begin{figure}[h!]
	\centering
	\includegraphics[scale=.70]{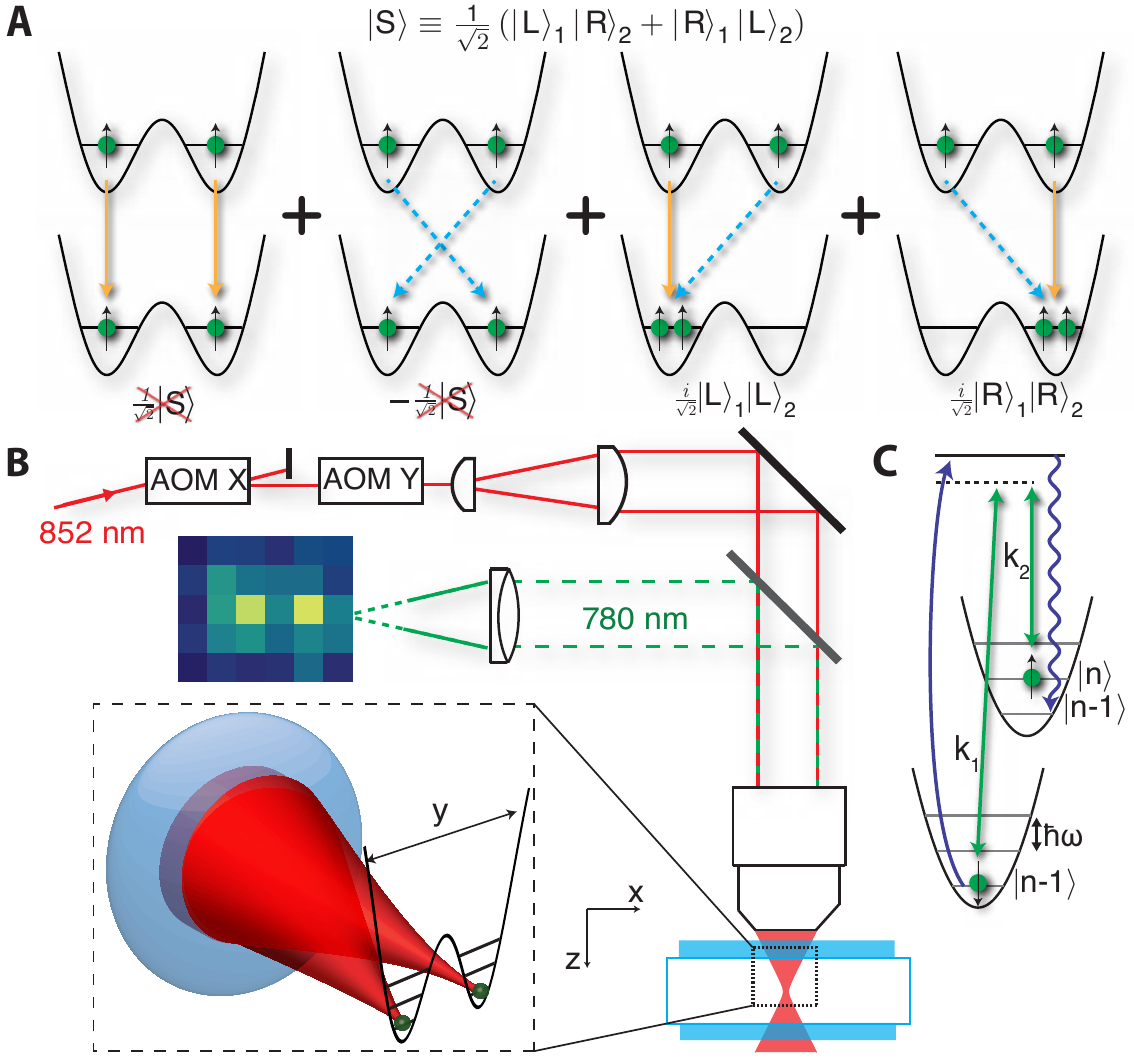}
	\caption{{\bf Hong-Ou-Mandel atom analog and experimental setup. (A)}  The optical tweezers form a coupled double-well potential. Starting from a state with a ground state spin-up atom in each well, denoted $\ket{S}$, the tunnel-coupling causes the atoms to interfere destructively and results in the state $\frac{i}{\sqrt{2}} ( \ket{L}_1\ket{L}_2+\ket{R}_1\ket{R}_2 )$. {\bf (B)} The apparatus for realizing tunneling between optical tweezers utilizes high numerical aperture optics combined with radio frequency signal control of the tweezers' positions and depths via acousto-optic modulators (AOMs). The same objective that creates the focused tweezer potentials also collects $780~\mathrm{nm}$ fluorescence from the optically trapped atoms. {\bf (C)} The sideband cooling is accomplished via lasers driving coherent (green) and spontaneous (blue) Raman transitions that couple to the atomic motion and spin states $\vert F=1, m_F = 1 \rangle \equiv \vert \!\!\downarrow \rangle$ and  $\vert F=2, m_F = 2 \rangle \equiv \vert \!\uparrow \rangle$.}
	\label{fig:exp}
\end{figure} 

The experimental results in this work depend on the mobility of two wavelength-scale optical tweezers and single-site imaging, which are realized via the apparatus illustrated in Fig.~1B~\cite{Schlosser2001,Jochim2011,Kaufman2012}.  For laser cooling to the three-dimensional (3D) ground state (Fig.~1C)~\cite{Monroe1995, Kaufman2012} and imaging in position-resolved potentials, our tweezers are positioned far apart compared to the focused spot radius of $710~\mathrm{nm}$.  For tunneling, the tweezers are brought close together such that there is a small, tunable overlap of the single-particle wavefunctions. The versatility of the tweezers, therefore, enables the simultaneous capabilities of laser-cooling in separated potentials and of coherently overlapping the single-particle wavefunctions with control on a scale much smaller than the size of the atomic wavepackets. Our full experimental sequence consists of the following steps: We image the initial atom positions, laser cool with Raman sideband cooling, perform tunneling experiments, and then image the atoms again.  Hence, we can follow the quantum dynamics between initial and final states that are both known with single-site resolution.

In Fig.~2A-C, we study the single-atom tunneling dynamics by only considering experiments that, after stochastic loading~\cite{Schlosser2001}, yield a single atom in the left or right well in the first image (fig.~S1).  After imaging and cooling, the atom is in the 3D motional ground state and the $\vert F=2, m_F = 2 \rangle \equiv \ket{\!\!\uparrow}$ spin state, where $F$ and $m_F$ are the total angular momentum quantum number and its projection along a quantization axis, respectively.  The tweezers are then dynamically reconfigured for tunneling experiments (Fig.~2A); both the depth and spacing are decreased rapidly (slowly) with respect to the tunnel-coupling (trap frequency) to prepare an initial state ideally localized in the left or right well (fig.~S2).   The tunnel-coupling is described by $J = -\int \phi_L(\vec{r}) H_{sp} \phi_R(\vec{r})d^3\vec{r}$, where $ \phi_L(\vec{r}) \equiv \bra{\vec{r}} L \rangle$ ($\phi_R(\vec{r}) \equiv \bra{\vec{r}} R \rangle$) is the lowest energy, localized wavefunction  for the left (right) well, and $H_{sp}$ is the single-particle Hamiltonian~\cite{Supplement}.  We control the energy bias $\Delta$ between the two wells by varying the relative intensity of each tweezer.  On the tunneling resonance ($\Delta = 0$), an atom prepared in the left well undergoes the coherent dynamics $\ket{L} \rightarrow \cos(Jt) \ket{L}  + i\sin(Jt) \ket{R}$~\cite{Folling2007,Anderlini2007,Strabley,Weitenberg2011}.  After an evolution time $t$ in the presence of tunneling, the depth of the traps is rapidly increased to freeze the atom distribution, the traps are pulled apart, and the single-atom location is imaged.   

\begin{figure*}[t!]
	\centering
	\includegraphics[scale =.435]{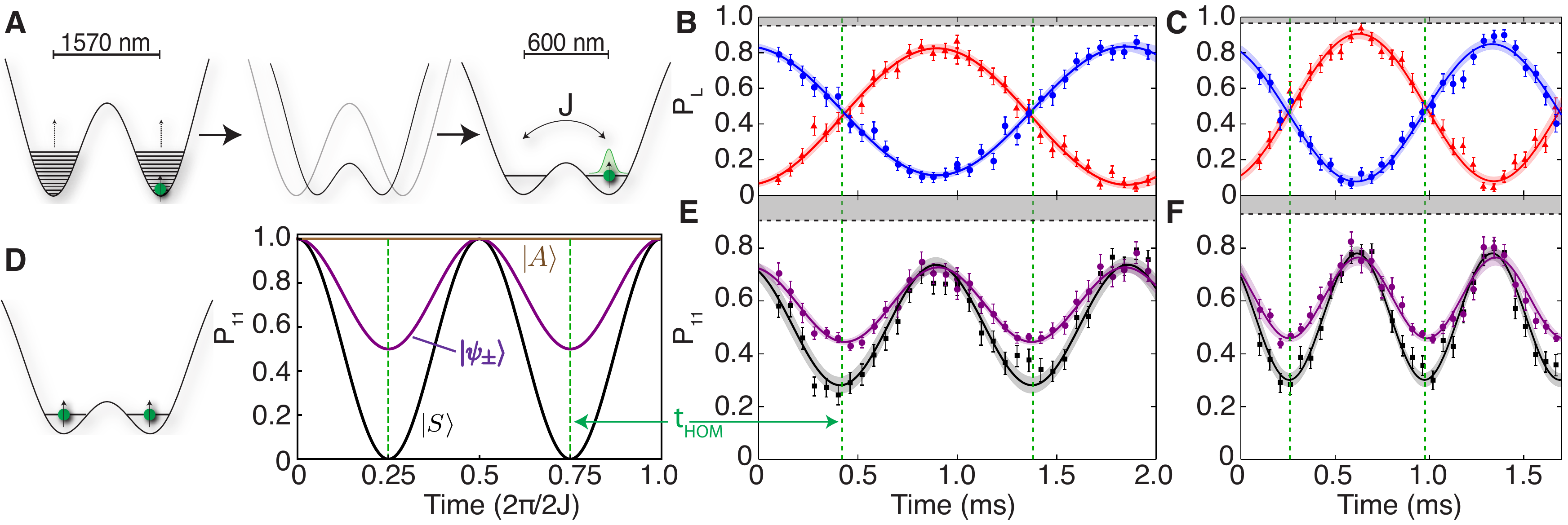}
	\caption{{\bf Single and two-particle tunneling.  (A)} Experimental overview. While the tweezers are $23(1)~\mathrm{MHz}$ deep, the atom is imaged, cooled and optically pumped to $\ket{\!\!\uparrow}$. For tunneling experiments, the tweezers are swept together such that the two gaussian functions are defined with an offset of $\approx800~\mathrm{nm}$ (resulting in double-well minima spaced by $\approx600~\mathrm{nm}$), and the total trap intensity is dropped by a large factor, resulting in a single-well depth of either 96~kHz or 60~kHz. {\bf (B)} Resonant tunneling oscillations at $2J$ for a $808~\mathrm{nm}$ gaussian function spacing and a 96~kHz depth.  Blue circles (red triangles) are the expectation value $P_{\rm L}^{1(2)}$ for finding an atom in the left well given an initial single atom in the left (right) well.  The gray shaded region indicates the contribution from atom loss $P_{\rm loss}$.  {\bf (C)} Same as B except with a  $805~\mathrm{nm}$ gaussian function spacing and a depth of 60 kHz.  {\bf (D)} Idealized two-particle tunneling dynamics.  Expectation for $P_{11}(t)$ for dynamics initiated at $t=0$ and in the symmetric spatial state $\ket{S}$, the distinguishable states $\ket{\psi_\pm}$, and the anti-symmetric state $\ket{A}$. The dashed green lines mark the locations of $t_{\rm HOM}$.  {\bf (E)}  Measured two-particle dynamics during the same experimental sequence as B.  Likelihood to measure exactly one atom in each well ($P_{11}$) for the initial condition in which an atom is prepared in each well (black squares).  Distinguishable expectation $P_{\rm dist}$ as determined from the single-particle data in B (purple circles).  The gray shaded region above the dashed black line indicates the expected reduction from atom loss.   {\bf (F)}  Same as E except here we realize a larger value of $J$ and smaller value of $U$ (see text) using the double-well parameters of C.  $t_{\rm HOM}$ for the experimental data is affected by a phase shift due to a small amount of tunneling before the nominal final trap is reached (fig S2); this effect is larger for faster tunneling.  In all plots, the shaded regions are the 95\% confidence interval for a sinusoidal fit.  The error bars are the standard error in the measurement; each black data point is the mean of $\approx140$ atom measurements, and each red or blue data point is the mean of $\approx100$ measurements.}
	\label{fig:exp}
\end{figure*} 

Figure 2B demonstrates resonant ($\Delta = 0$) coherent tunneling as measured by recording the likelihood of observing the atom in the left well ($P_{\rm L}$) as a function of time for an atom starting in the left well (blue) or the right well (red).  The tunneling can be tuned in accordance with our expectations by varying either the tweezer spacing or the overall potential depth (fig.~S3C).  In Fig.~2B, a fit to the data reveals $J/2\pi=262(4)$ Hz; Fig.~2C shows data in which $J$ is increased to $348(4)$ Hz.  Part of the finite contrast of the oscillations is due to atom loss due to background collisions; in the duration of our experiments the loss probability ($P_{\rm loss}$) ranges from 0.03 to 0.05 and is known precisely for each experiment (fig.~S1, table S1) (Figs.~2B,C gray regions).  We also observe damping of the tunneling oscillations and finite initial contrast that is not accounted for by particle loss ($\tau\approx 10$ ms for $J/2\pi=262$ Hz) (fig.~S3B).  These effects are most likely due to experimental fluctuations of the double-well bias, and we experimentally find that the contrast and damping improve with increasing $J$~\cite{Supplement}.

We now consider the theoretical expectation for an equivalent dynamical experiment starting with two particles, one atom in each well.   For perfect cooling and spin preparation of the isolated atoms, all degrees of freedom besides their position (left or right) will have been made the same, i.e.~we know there is a particle in the left well and there is a particle in the right well, but we cannot associate any additional label to the particles.  The bosonic atoms will then, necessarily, occupy the spatially symmetric $\ket{S}$ state.  For poor cooling or spin preparation, the atoms can be distinguished by a degree of freedom other than their position; hence, the atoms can anti-symmetrize in the additional degree of freedom, motional state or spin, and in turn have a projection onto the anti-symmetric spatial state $\ket{A} \equiv \frac{1}{\sqrt{2}} (\ket{L}_1\ket{R}_2 -\ket{R}_1\ket{L}_2 )$. The bosonic state can then be written as a mixture of the states $\ket{\psi_{\pm}} = \frac{1}{\sqrt{2}} (\ket{S}\ket{\chi_+} \pm \ket{A}\ket{\chi_-})$, where $\ket{\chi_{\pm}} = \frac{1}{\sqrt{2}} (\ket{\chi}_1\ket{\bar{\chi}}_2\pm\ket{\bar{\chi}}_1\ket{\chi}_2 )$ and $\{\chi,\bar{\chi}\}$ describe the other degree of freedom such as motional state $\{n, n'\}$ or spin $\{\uparrow, \downarrow\}$. Two atoms in either of the $\ket{\psi_{\pm}}$ states are distinguishable because the additional degree of freedom $\{\chi,\bar{\chi}\}$ is uniquely correlated with the atoms' positions: For the $\ket{\psi_+}$ ($\ket{\psi_-}$) state, the atom on the left is in state $\ket{\chi}$ ($\ket{\bar{\chi}}$) and the atom on the right is in state $\ket{\bar{\chi}}$ ($\ket{\chi}$).  The ability to measure two-atom indistinguishability arises from the different dynamics exhibited in the symmetric and anti-symmetric cases. The symmetric spatial state dynamically evolves as $\ket{S} \rightarrow \ket{S} \cos(2Jt)+\frac{i}{\sqrt{2}} (\ket{L}_1\ket{L}_2+\ket{R}_1\ket{R}_2)\sin(2Jt)$. The anti-symmetric state $\ket{A}$ undergoes destructive interference that prevents the two bosons from being in the same well, and hence displays no tunneling dynamics.

In Fig.~2D we show the expected {\it ideal} dynamics for the distinguishable (purple) and indistinguishable (black) cases; we emphasize that $t$ is the time the atom spends tunneling.  We consider the observable $P_{11}(t)$, which is the likelihood to measure the atoms in separate wells as a function of time spent tunneling.  A measurement of $P_{11}$ is analogous to looking at coincidence counts on a pair of photon detectors.  For the distinguishable states $\ket{\psi_{\pm}}$,  $P_{11}(t)$ is the average of the time-independent $P_{11}(t) = 1$ from $\ket{A}$ and unity contrast oscillations of $P_{11}(t)$ from $\ket{S}$ (Fig.~2D), and as such does not attain a value below 0.5.   The $\ket{S}$ state oscillates with unity contrast and thus vanishes at times $t_{\rm HOM}=2\pi/8 J$ and odd multiples thereof (green dashed lines).  At $t_{\rm HOM}$, in analogy to the HOM effect for photons, each atom has been coherently ÒbeamsplitÓ between the two wells.

In Fig.~2E, we experimentally investigate the population dynamics observed with two particles.  We plot $P_{11}(t)$ for cases in which the stochastic loading results in two atoms, one in each well (black squares); these points are taken in the same experimental sequence as the single-particle data in Fig.~2B.  In our atom detection protocol, we image scattered light from the two well-separated traps onto a CCD array.  During the 25 ms to 50 ms imaging time, the atoms are cooled via polarization gradient cooling, and due to light-assisted atomic collisions we observe signal corresponding to either zero or one atom~\cite{Schlosser2001,Bakr2009,Weitenberg2011}.  $P_{11}$ is determined by the distinct signature in which the image indicates one atom in each well.  If the experiment yields two atoms in one well, $P_{20}$ or $P_{02}$, this is manifest by final images that yield zero atoms, or in some cases one atom in a single well (fig.~S1)~\cite{Supplement}.  To accurately interpret $P_{11}$ we take into account signal depletion due to the single-particle loss described earlier ($P_{\rm loss}$).  This effect reduces the maximum value that can be achieved by the measured $P_{11}$ to $(1-P_{\rm loss})^2$.  (The gray region above the black dashed line in Figs.~2E,F indicates the loss contribution).

Before focusing our studies on experiments at $t_{\rm HOM}$, we first analyze the result of the full two-particle dynamics.  Our goal is to compare $P_{11}(t)$ from our two-particle measurement to that of a theoretical expectation for uncorrelated, distinguishable atoms, which we refer to as $P_{\rm dist}(t)$.  $P_{\rm dist}$ at any time can be calculated directly from corresponding single-particle data via $P_{\rm dist}=P_{\rm L}^1P_{\rm R}^2+P_{\rm R}^1P_{\rm L}^2$ (purple circles in Fig.~2E)~\cite{Supplement}.  Here, $P_{\rm L}^{1(2)}$ corresponds to measuring an atom in the left well when an atom starts in the left (right) well, i.e.~the blue (red) data of Fig.~2B, and $P_{\rm R}^{1(2)}$ is the corresponding information for measuring an atom in the right well.  A calculation of $P_{\rm dist}$ directly from the single-particle points inherently contains both loss and finite single-particle contrast.  For example, $P_{\rm dist}(t_{\rm HOM})$ reaches a value consistent with $(1-P_{\rm loss})^2/2\approx0.5-P_{\rm loss}$, and the amplitude of $P_{\rm dist}(t)$ is consistent with the expectation of one half the product of the single-particle contrasts~\cite{Supplement}. We can compare the amplitude of oscillation for the distinguishable expectation (purple circles) to our two-particle measurement (black squares).  We find these values differ by $6\sigma$~\cite{Supplement}:  $A_{P_{\rm dist}}=0.282(12)$ and $A_{P_{11}}=0.46(2)$.  

A full treatment of the observed $P_{11}(t)$ must also consider potential effects of interactions between the atoms.  In many experiments with atoms in optical lattices the on-site interaction energy $U$ is the dominant scale~\cite{Folling2007,Ma2011}; however, we intentionally operate in a regime where $U$ is smaller than $J$.  For the data shown in Fig.~2E, $U=0.44(4)J$~\cite{Supplement}.  In Fig.~2F we demonstrate a similar HOM signature for experimental conditions of even smaller relative interaction $U=0.22(2)J$, with measurements $A_{P_{11}}=0.48(2)$ and $A_{P_{\rm dist}}=0.306(18)$.  The similarity of these results to those in Fig.~2E suggests that interactions are not a relevant scale in either experiment.  Nevertheless, we also theoretically analyze whether interactions between distinguishable atoms could mimic the HOM signal~\cite{Supplement} (figs.~S5,S6).  For the two-particle initial conditions expected in our experiment, the interaction energy shift suppresses two-particle tunneling regardless of distinguishability, and hence the theoretical expectation for $A_{P_{\rm dist}}$ decreases with increasing interactions. In the implausible case that the particles initially have specific coherences described in~\cite{Supplement}, $A_{P_{\rm dist}}$ can be larger and $P_{\rm dist}(t)$ can reach a lower minimum value. Even in this unlikely circumstance our data are statistically different from this interacting distinguishable case (fig.~S6).  Another independent piece of evidence that the HOM interference we observe arises due to quantum statistics and not atom interactions comes from the studies presented below.  In these experiments we vary the two-particle spin state and consequently the HOM interference, while changing the interactions by at most $5\%$~\cite{Egorov2013}.

We now focus on experiments in which we fix the tunneling time at $t_{\rm HOM}$, where $P_{11}$ reaches a value that we refer to as $P_{\rm HOM}$.  By studying the behavior of $P_{\rm HOM}$ as the indistinguishability of the atoms is varied, we can observe the analog of an ``HOM dip".  In particular, we vary the two-particle spin state and the 3D motional ground-state fraction to introduce distinguishing spin or motional degrees of freedom.  

\begin{figure}[h!]
	\centering
	\includegraphics[scale=0.32]{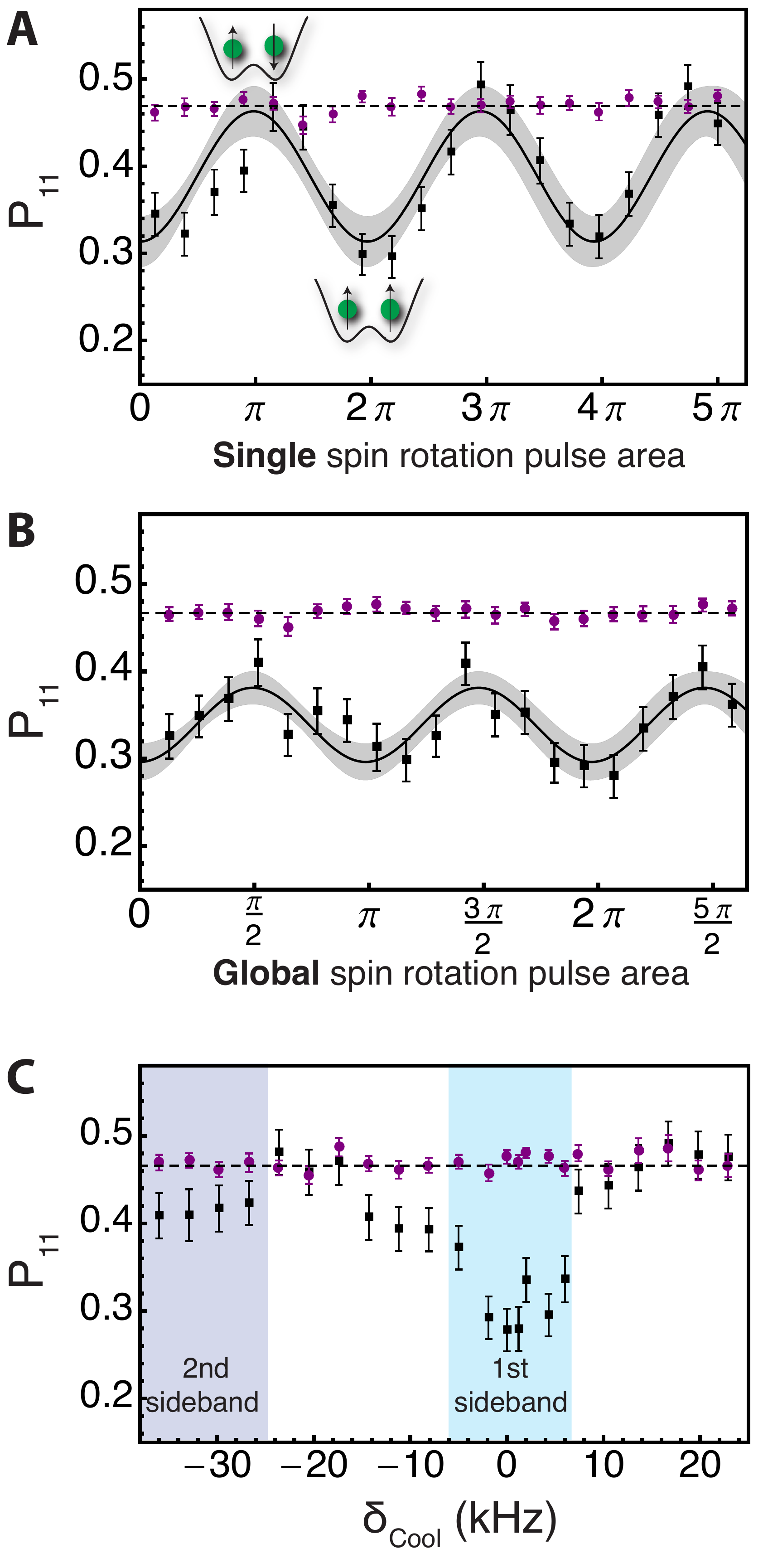}
	\caption{{\bf The HOM effect observed by varying atom distinguishability.}  In all plots the black squares are $P_{11}(t_{\rm HOM})$=$P_{\rm HOM}$, the purple circles are the expectation for distinguishable particles calculated directly from the single-atom tunneling ($P_{\rm dist}(t_{\rm HOM})$), and the dashed black line marks $(1-P_{\rm loss})^2/2$. {\bf (A)} Before tunneling we apply a microwave drive that couples $\ket{\!\!\uparrow}$ and $\ket{\!\!\downarrow}$ for one of the atoms in a two-particle experiment.  In the trap where $J/2\pi=348$ Hz the tunneling time is fixed at $t= 0.99~\mathrm{ms}$ (second realization of $t_{\rm HOM}$).  {\bf (B)} Before tunneling we apply a global coherent drive of varied pulse area to couple $\ket{\!\!\uparrow}$ and $\ket{\!\!\downarrow}$ and then allow for decoherence. In the trap where $J/2\pi=262$ Hz the tunneling time is fixed at $t= 0.45~\mathrm{ms}$.   In A and B the solid line and shaded band are sinusoidal fits and the associated 95\% confidence interval.  {\bf (C)} HOM dip dependence on cooling.  We vary the detuning ($\delta_{\rm Cool}$) of the cooling beams of motion along the $z$-axis. In the trap where $J=262$ Hz the tunneling time is fixed at $t= 0.45~\mathrm{ms}$.  The two shaded regions correspond to frequency ranges of efficient (1st sideband) and less efficient (2nd sideband) cooling. For all plots, each black data point is the average of $\approx360$ measurements, and each set of measurements corresponding to a purple point is the average of $\approx240$ measurements.  All error bars are the standard error in the measurement.}
	\label{fig:exp}
\end{figure}

We start by studying the dependence of $P_{\rm HOM}$ on the relative spin state of the two atoms using two distinct methods.  In the first method, after cooling the atoms, we apply a variable-length microwave pulse that couples the $\ket{\!\!\uparrow}$ and $\ket{\!\!\downarrow}$ spin states in only the right well (fig.~S4).  This is accomplished by shifting the transition in the left well out of resonance using a circularly-polarized, tightly focused laser spot~\cite{Supplement}.  Upon $\pi$ rotation of this spin, the atoms become distinguishable, and we expect the HOM dip to disappear.  The observed dependence on the microwave pulse area is shown in Fig.~3A; for comparison we show that $P_{\rm dist}$, from the single-particle measurements, remains constant (purple circles).   We study multiple spin rotations to show that the HOM effect is recovered after a $2\pi$ rotation.  We find the frequency of oscillation is $32.6(6)~\mathrm{kHz}$, which is in agreement with the measured microwave Rabi frequency of $32.05(18)~\mathrm{kHz}$.  The displayed fit to the data determines $P^{\rm min}_{\rm HOM}=0.314(14)$, and the amplitude of the variation in $P_{\rm HOM}$ is $0.15(2)$.  Taking into account the spin rotation fidelity we expect an amplitude of $0.84((1-P_{\rm loss})^2/2-P^{\rm min}_{\rm HOM})=0.130(13)$~\cite{Supplement}, which is consistent with the measured value.

In the second spin study, we simultaneously couple the $\ket{\!\!\uparrow}$ and $\ket{\!\!\downarrow}$ spin states of atoms in both wells using a pair of Raman beams.  This global rotation avoids any systematic effects that might be introduced by single-site addressing.  During the time (25 ms) between the Raman pulse and the tunneling, the atoms lose their spin coherence, and hence the spin state of each atom is in an incoherent mixture.  Ideally, odd multiples of a $\pi/2$-pulse yield an equal mixture of all possible two-atom spin states, and hence the likelihood for the atoms to have opposite spin is $1/2$ and the magnitude of the HOM dip should be reduced by $1/2$. The observed dependence on the Raman pulse area is shown in Fig.~3B.   We find the frequency of oscillation is $65.5 \pm 1.2~\mathrm{kHz}$, which, as expected, is {\it twice} the measured Raman Rabi frequency of $32.3(3)~\mathrm{kHz}$.  The displayed fit to the data determines $P^{\rm min}_{\rm HOM}=0.296(10)$, and the amplitude of the variation in $P_{\rm HOM}$ is $0.085(15)$.  Taking into account the relative spin rotation fidelity of $0.90(3)$, we expect a half amplitude of $(0.90^2/2)((1-P_{\rm loss})^2/2-P^{\rm min}_{\rm HOM})=0.069(6)$, which is consistent with the measured amplitude.

Lastly, we study the dependence of $P_{\rm HOM}$ on the motional state of the atoms in Fig.~3C.  During the last stage of our cooling we vary the frequency $\delta_{\rm Cool}$ of one Raman beam that controls the cooling along the weak axis ($z$) of both tweezer wells.  For a separable potential, motional excitation along this axis would leave the single-particle tunneling unaffected.  For our non-separable tweezer potential, we expect and observe some variation in the tunneling~\cite{Supplement}, but near $t_{\rm HOM}$ the single-particle tunneling still results in a relatively constant $P_{\rm dist}$ (purple circles), which is consistent with the distinguishable expectation $(1-P_{\rm loss})^2/2=0.4660(14)$.   For the two-particle measurements, at the primary sideband cooling resonance ($\delta_{\rm Cool}=0$), we observe a dip to $P^{\rm min}_{\rm HOM} = 0.28(2)$, which is a value below the distinguishable expectation.

In all of the measurements we present, the value of $P^{\rm min}_{\rm HOM}$ is observed to be finite, and it is useful to consider the origin of imperfections that could lead to the observed value.  As an estimate of one effect, if residual atom temperature along the weak $z$-axis were the only contributing factor, the central value of our single-atom ground state fraction (measured via sideband spectroscopy to be $85\%$~\cite{Supplement,Kaufman2012}) would correspond to $P_{\rm HOM}^{\rm min} = 0.12$.   Hence, it is likely that the finite $P_{\rm HOM}^{\rm min}$ is further enlarged by technical fluctuations similar to those that lead to our finite single-particle contrast.

We have demonstrated a new experimental system with which we achieve quantum control over the motion, position, and spin of single neutral atoms. Through an HOM experiment with massive composite bosons, we have shown that with laser cooling alone it is possible to generate a low-entropy set of atoms in optical tweezers, where quantum statistics and atomic indistinguishability are manifest.  Our work introduces a new platform for quantum-state engineering and for studying the interplay between spin and motional degrees of freedom that is key to many condensed matter models. 

We thank T. P. Purdy for useful discussions.  This work was supported by the David and Lucile Packard Foundation and the National Science Foundation under grant number 1125844.  CAR acknowledges support from the Clare Boothe Luce Foundation, AMK and CMR from NDSEG, and BJL from NSF-GRFP. KRAH, MLW, AMR acknowledge funding from NSF-PIF, ARO, ARO-DARPA-OLE, and AFOSR. KRAH and MFF acknowledge support from the NRC postdoctoral fellowship program. 

%\bibliographystyle{Science}
%\bibliography{atomrefs2.bib}

\begin{widetext}

\newcommand{\cumulant}[1]{\ev{#1}_c}
\newcommand{\ev}[1]{\langle #1 \rangle}
\newcommand{\tr}[1]{ {\rm tr}[#1]}
\newcommand{\comment}[1]{{\tt (#1)}}
\newcommand{\todo}[1]{ {\red\bf (#1)}}
\newcommand{\ddt}[1]{\frac{d #1}{dt}}
\newcommand{\dby}[1]{\frac{\partial}{\partial #1}}
\newcommand {\micro}[1]{$\mu$#1}
\newcommand {\K}{\mathbf{k}}
\newcommand{\mathhuge}[1]{\mathlarger{\mathlarger{\mathlarger{\mathlarger{\mathlarger{#1}}}}}}

% The preamble here sets up a lot of new/revised commands and
% environments.  It's annoying, but please do *not* try to strip these
% out into a separate .sty file (which could lead to the loss of some
% information when we convert the file to other formats).  Instead, keep
% them in the preamble of your main LaTeX source file.

% The following parameters seem to provide a reasonable page setup.

%The next command sets up an environment for the abstract to your paper.

\newenvironment{sciabstract}{%
\begin{quote} \bf}
{\end{quote}}

% If your reference list includes text notes as well as references,
% include the following line; otherwise, comment it out.

\renewcommand\refname{References and Notes}
\newcommand{\lp}{\ensuremath{\left(}}
\newcommand{\rp}{\ensuremath{\right)}}
\newcommand{\cred}{\color{red}}
\newcommand{\cblack}{\color{black}}
\newcommand{\e}[1]{\ensuremath{\times 10^{#1}}}

% The following lines set up an environment for the last note in the
% reference list, which commonly includes acknowledgments of funding,
% help, etc.  It's intended for users of BibTeX or the {thebibliography}
% environment.  Users who are hand-coding their references at the end
% using a list environment such as {enumerate} can simply add another
% item at the end, and it will be numbered automatically.

\newcounter{lastnote}
\newenvironment{scilastnote}{%
\setcounter{lastnote}{\value{enumiv}}%
\addtocounter{lastnote}{+1}%
\begin{list}%
{\arabic{lastnote}.}
{\setlength{\leftmargin}{.22in}}
{\setlength{\labelsep}{.5em}}}
{\end{list}}

\setcounter{figure}{0}
\renewcommand{\theequation}{S\arabic{equation}}
\renewcommand{\thefigure}{S\arabic{figure}}
\renewcommand{\thetable}{S\arabic{table}}

\

\

\begin{center}
{\huge Supplementary Materials}
\end{center}

\

\noindent{\Large This section includes:}
\begin{itemize}
\item Materials and Methods
\item Figures S1-S6, Table S1
\end{itemize}

\

\noindent {\bf \Large Materials and Methods}
 
\section{Experimental protocols, calibration, and supporting data}
\label{experiment}

\subsection{The optical tweezer potential}
\label{sec:tweezer}

The optical tweezer potentials used for single-atom trapping are generated by focusing 852 nm light through a high numerical aperture (NA) lens.  The system aperture is set to operate at 0.6 NA, and the nominally gaussian input beam is clipped by the aperture at 0.88 of the gaussian waist ($1/e^2$ radius). The finite aperture alters the ideal focus from a gaussian profile, and due to imperfections in the optical system we know that there are also some small aberrations that further reduce the trap strength. Here we describe the methods used to develop a quantitative understanding of our trapping potential.

We can measure the trap depth by recording the light shift of the $| F=2 , m_F=2 \rangle \rightarrow |F^\prime=3,m_F^\prime=3\rangle$  cycling transition of $^{87}$Rb as a function of the input power.   The resonance is recorded at each trap depth by scanning the frequency of a $\sigma^+$-polarized probe beam, which heats the atom out of the trap when on resonance with the cycling transition. Because the shift of the cycling transition is primarily determined by  the peak intensity at the tweezer focus ($I_0$), we can extract the peak intensity as a function of input power \cite{RauschenbeutelLightShift}. From $I_0$ we can directly calculate the trap depth ($V_0$).  We also measure directly the trap oscillator frequencies~\cite{Kaufman2012Supp}, which determine the curvature at the bottom of the tweezer potential.   

To determine our best estimate for the overall intensity profile we calculate the expected aberration-free Huygens point spread function (PSF) in Zemax.  Due to the clipped gaussian input beam, this results in a slightly non-gaussian function.  Matching this function to the measured trap depth and curvature provides the tweezer potential; to simplify our tunneling calculations we then fit the full potential to a gaussian to parameterize the trap in terms of a depth, and effective waist ($w_0$) and independent $z$-length scale ($z_0$).  We find a waist of $710(10)$ nm, and for the input power used during the cooling and imaging the depth of each trap is 23(1) MHz.  The corresponding trap depths during the tunneling dynamics are 96(4) kHz and 60(4) kHz.

\subsection{Calibration of tweezer trap spacing}
\label{sec:spacing}

The tunneling is extremely sensitive to the distance between the optical tweezers. To calibrate the spacing we measure the imaging magnification of the optical system in a separate setup using a custom fabricated test target with two approximately 348 nm diameter pinholes separated by 6861 nm.  From this test setup, we measure an imaging magnification of $M=48.9(5)$.  We then image the atoms on our CCD array as a function of the frequency difference of the RF tones that generate the optical tweezers. Via the magnification we then find the gaussian function spacing of the optical tweezers is given by 0.209(3) $\mu\text{m}/\text{MHz}$.   Note in our experiments the gaussian function spacing of $\approx 800$ nm ($a$ in Eq.~(\ref{eq:GaussPot})) is similar to the trap waist; in this situation the spacing between the double-well minima is different (smaller) than the functional spacing between the gaussians. 

\subsection{Determining $P_{11}(t)$ from imaging data}
\label{sec:imaging}

In the experimental two-particle data presented in the main text we focus on the quantity $P_{11}(t)$.   Here we discuss how we measure $P_{11}$ and its relation to $P_{20}$ and $P_{02}$.  In our experiments we start by stochastically loading atoms into each tweezer.  Due to light-assisted collisions we only ever observe zero or one atoms in the trap after our loading sequence~\cite{Schlosser2001Supp}, and for our trap and cooling parameters we find we achieve approximately $60\%$ loading of a single atom.  A key feature of our experiment is that we have a record of the initial number of atoms in each well for every experiment because we image before performing Raman cooling and initiating the tunneling experiment.  In each iteration of the experiment we then take a final image after tunneling.  As discussed in the main text, during the imaging time, the atoms are cooled via polarization gradient cooling, and due to light-assisted atomic collisions we observe signal corresponding to either zero or one atom.  Figure~\ref{fig:imaging} illustrates the full set of possible outcomes of the two sets of images for a double well; it also provides a visual picture of how both single-particle and two-particle experiments arise from the same experimental sequence.  

$P_{11}$ is defined by the case in which both images indicate one atom in each well.  If the experiment yields two atoms in one well, $P_{20}$ or $P_{02}$, this is manifest by final images that yield zero atoms, or in some cases one atom in a single well (a ``two-to-one event").  More explicitly, in our calculation of $P_{11}$ if there is an atom in each well we count this as 1; if there are zero atoms, we count this as 0; if there is one atom total, we also count this as 0.  We then take the mean over all experimental realizations.  To accurately interpret the measured $P_{11}$ we must take into account particle loss.  Hence, in our analysis this loss is independently accounted for by using the value of $P_{\rm loss}$ determined in the parallel single-particle experiments.  Specifically, in two-particle experiments the maximum value that $P_{11}$ can reach is $(1-P_{\rm loss})^2$.  For the small values of $P_{\rm loss}$ observed in our experiments to good approximation this is $1-2P_{\rm loss}$.  $P_{\rm loss}$ ranges in our experiments between 0.03 and 0.05; these values are consistent with variation in vacuum lifetime and experiment length amongst different datasets.  

Independent of this procedure, we can study the two-to-one events we observe and confirm that they are a signature of two atoms on a single well.  This perspective affirms our treatment above, but note our analysis does not directly depend upon the fraction of images that yield two-to-one events versus two-to-zero events.  First, we find that in our experiments we see an increase in two-to-one events when $P_{11}(t)$ is minimal, i.e.~when the likelihood of finding two atoms on the same tweezer is maximal.  Directly from the data shown in Fig.~2E and the calibrated single-atom loss we can extract the ratio of two-to-one and two-to-zero events and find 29(4)\% of the time a two-to-one event occurs when the data is analyzed at $t_{\rm HOM}$ (minimum of $P_{11}$) and a consistent value of 22(5)\% when the data is analyzed at the maximum of $P_{11}$.  We also can directly measure two-to-one events by carrying out a separate experiment in which we combine two traps each with a single atom to deterministically start with two atoms in a single trap. In this experiment we find 26(2)\% of the time a two-to-one event occurs.  These findings are in contrast to many optical lattice experiments in which pure parity imaging is observed~\cite{Bakr2009Supp, Weitenberg2011Supp}. However, in other optical tweezers experiments in which atoms undergo light-assisted collisions it has also been observed that images of two atoms do not yield zero or one atom based on the parity of the atom number~\cite{Sompet2013}.

\subsection{Single-particle preparation: 3D sideband cooling and  thermometry}
\label{sec:cooling}

After the first imaging of the initial atom configuration, we perform 3D Raman-sideband cooling.  Our 3D cooling procedure uses the same beam geometry as our earlier work~\cite{Kaufman2012Supp}. However, we modified our cooling protocol so that it operates in a semi-continuous, as opposed to a pulsed, format. The repumping ($F=1-2'$) and optical pumping beams ($F=2-2'$) are on for the entirety of the cooling procedure, during which their powers are modulated depending on the relevant motional axis being addressed. The modification is a trade off between cooling time (nearly 5 times longer than the pulsed approach in our previous work) and stability. We find this approach is much more stable to drifts in the Raman Rabi frequencies and pumping scattering rates, especially along the weakly confining axial ($z$) axis.

There are 100 cooling cycles, and within each cycle we alternate between the axial and then radial trap dimensions. For
the first 75 cycles, we cool on the first (second) sideband for the radial (axial) direction, while for the last 25 cycles we cool on the first sideband for both dimensions.  As in our previous work, the radial sidebands indicate a lower bound on our radial ground state fraction of $P^{\rm GS}_{\rm radial} \gtrsim 95\%$. The limitation to our 3D ground state fraction is largely determined by our residual axial motion.  Based on spectroscopy taken directly after the data in Fig.~2B,E, we conclude an axial ground-state fraction of of $P^{\rm GS}_{\rm axial} = 85^{+12}_{-10}\%$. Hence, we expect a 3D ground-state fraction of approximately $80\%$.

For the data displayed in Fig.~3C, it is during the last 25 cycles that we adjust the detuning of the Raman beams that address the axial dimension. While the first 75 cycles realize an axial ground-state fraction of $P^{\rm GS}_{\rm axial} \approx 50\%$, the final ground-state fraction depends on the last 25 cycles of first sideband cooling. Hence, by varying the cooling detuning at this stage, we vary the axial ground-state fraction between $P^{\rm GS}_{\rm axial} \lesssim 50\%$ and the final value $P^{\rm GS}_{\rm axial} = 85^{+12}_{-10}\%$. 

\subsection{Single-particle preparation: Tunneling}
\label{sec:preparation}

Once 3D cooling has finished, the tunneling experiments start. The preparation proceeds as follows and is illustrated in Fig.~\ref{fig:tunneling_prep}A.  (All numbers presented here are for the final trap values that realize the smaller value of $J$ presented in Fig.~2B,E and Fig.~3B,C.) While the traps are each still $23(1)~\mathrm{MHz}$ deep, the gaussian tweezer spacing is swept from $1570~\rm nm$ to $808~\rm nm$ in $T_{\rm pos} = 10~\mathrm{ms}$. Afterward, in a $T_1=10~\mathrm{ms}$ linear ramp the traps are dropped to $2.3(1)~\mathrm{MHz}$, and from there, in a second linear ramp of $T_2 = 5~\mathrm{ms}$ to a depth of $96(4)~\mathrm{kHz}$. This last ramp initiates the tunneling experiments. After a variable evolve time $t$, the trap is jumped in $100~\mathrm{\mu s}$ to $210~\mathrm{kHz}$ (not shown in Fig.~\ref{fig:tunneling_prep}A), and then linearly ramped in $T_3 = 10~\mathrm{ms}$ back to $23(1)~\mathrm{MHz}$. The traps are then swept apart in $T_{\rm pos} = 10~\mathrm{ms}$ to $1570~\mathrm{\rm nm}$ for imaging. The same procedure is used for experiments in which the final tunneling depth is 60 kHz and the well spacing is 805 nm (Fig.~2C,F and Fig.~3A). 

The goal of the second ramp $T_2$  is to minimize tunneling that occurs before $t=0$, while preserving the temperature achieved during sideband cooling. The data in Fig.~\ref{fig:tunneling_prep}B affirms that minimal, though measurable, population transfer occurs before our nominal $t=0$. Based on the data in Fig.~2B, we observe that the coherent dynamics are consistent with a small time offset $t_{\rm corr} = -60~\mu s$, which informs our calculation of the experimental $t_{\rm HOM} =  2\pi/8J+t_{\rm corr} =0.42~\rm ms$ for $2J/2\pi = 524~$Hz.   With respect to any heating during the ramps, if we do the tunneling initiation procedure forward and then in reverse, we do not observe changes in the temperature measured via sideband spectroscopy. Lastly, we note that the data in Fig.~\ref{fig:tunneling_prep}B validate assumptions in the theoretical section below that interactions during the ramp can be neglected, because very little population transfer occurs in this time.

\subsection{Single-particle tunneling dependences}
\label{sec:tunneling}

Here we present additional data on single-particle tunneling in a double-well potential.  These studies help elucidate the origin of the single-particle tunneling contrast and damping, and they are verification of our understanding of the optical tweezer potential parameters (Sec.~\ref{sec:tweezer}) that are later important for determining the on-site interaction energy (Sec.~\ref{sec:interaction}).  
 
In Fig.~\ref{fig:tunneling}A,B we display single-particle tunneling data for a double-well trap with a final single-well depth of 96 kHz and an $a=808$ nm gaussian function spacing.  In Fig.~\ref{fig:tunneling}A  we plot the likelihood of observing the atom in the left well ($P_{\rm L}$) as a function of the double-well bias $\Delta$ for a fixed evolution time of 0.9 ms for the condition of an atom initially imaged on the left (blue) or on the right (red).  We estimate a 15$\%$ uncertainty in $\Delta$.  In Fig.~\ref{fig:tunneling}B we plot tunneling oscillations out to longer times than displayed in similar data in Fig.~2B; in this trap we find a damping $\tau \approx 10$ ms.  We also demonstrate in Fig.~\ref{fig:tunneling}C a tunability unique to our apparatus by measuring the tunneling rate as a function of the gaussian function spacing $a$ (while maintaing a 96 kHz single well depth).  We observe in accordance with theory that the tunneling changes rapidly with the spacing.  The solid line is a theoretical expectation (see Sec.~\ref{sec:DVR}) for the trend calculated for an ideal gaussian beam of waist 707 nm and a single-well depth of 96~kHz.  These values are consistent with our independently determined trap and spacing parameters (Sec.~\ref{sec:tweezer} and Sec.~\ref{sec:spacing}), indicating our absolute knowledge of the trap parameters.  We have also studied similar single-particle tunneling for a range of $J/2\pi$ values from 150 to 350 Hz.  We find that the initial contrast decreases and the damping increases with smaller $J$.  This behavior is consistent with the presence of technical fluctuations in the trap; in particular our models point to fluctuations in the double-well bias.

The tunneling calculation shown in Fig.~\ref{fig:tunneling}C assumes ground-state tunneling.  We have also analyzed theoretically tunneling rates in higher vibrational states using the calculation presented in Sec.~\ref{sec:DVR}.  For our experiments the main contribution to $J$ variations is predominantly-axial excitations.  In a separable potential, motional excitation along the axial $z$-axis would leave the single-particle tunneling in $y$ unaffected. For our non-separable tweezer potential we expect some variation in $J$.  For typical trap parameters $J$ is changed by $\approx40\%$ over the first three vibrational states.  In the data presented in Fig.~3C, where the axial excitation is varied, we observe single-particle data consistent with fluctuations in the tunneling.  In addition at the highest values of $\delta_{\rm cool}$ we observe some asymmetry between the wells.  However at $t_{\rm HOM}$, $P_{\rm dist}$ is first-order insensitive to the exact value of $J$, and we find experimentally that $P_{\rm dist}$ is relatively unchanged over the relevant range of cooling (purple circles).  We also expect variations in $J$ due to axial excitation to introduce dephasing in single-particle tunneling oscillation data.  In our current experiments, however, we estimate the dephasing is dominated by technical fluctuations discussed above and not axial cooling fidelity.

\subsection{On-site interaction energy}
\label{sec:interaction}

In our experiment we determine the on-site interaction energy by using independent knowledge of our trap parameters (well depths, gaussian center spacing, and the effective gaussian waist) as inputs to the theoretical calculations described in Sec.~\ref{sec:DVR}. In performing these calculations of $J$ and $U$, we assume that the tweezer potentials are gaussian and identical.  Propagating the uncertainties in the trap parameters, as described in Sec.~\ref{sec:tweezer} and Sec.~\ref{sec:spacing}, we find the uncertainty in the predicted $U$.  Note due to the extreme sensitivity of $J$ to trap parameters we can predict $U$ with less uncertainty than an equivalent prediction of $J$.  The values of $U/J$ presented in the manuscript use the measured $J$ (and associated statistical uncertainty from the oscillation measurement) and calculated $U$ to generate our best estimate of $U/J$ in each of the traps.

\subsection{Spin-flip of a single atom in the double well}
\label{sec:spinprotocols}

We achieve the single-atom spin addressing in Fig.~3A by focusing a ``light-shift" beam on the left trap when the trap spacing is $1570$ nm and the traps are still at their largest depth, while simultaneously irradiating both traps with a microwave signal.  The goal is to achieve a significant shift of the $\ket{2,2}\leftrightarrow\ket{1,1}$ microwave transition for the left atom only.  The circularly-polarized light-shift beam is $-50$~GHz detuned from the D2 line and contains $250\rm~nW$ of power in a beam with a $\approx 0.7~\mu \mathrm{m}$ radius.   We observe that on a 10-hour time scale the light shift beam can drift off the left trap, and hence we regularly recalibrate its position and measure the microwave response of each well. In Fig.~\ref{fig:spinFlip}, we show data demonstrating this technique, recorded directly following the data shown in Fig.~3A. Figure~\ref{fig:spinFlip}A demonstrates that the resulting relative shift of the microwave transition between the left well and right well is nearly $200~\rm kHz$. In Fig.~\ref{fig:spinFlip}B, we show the transition probability as a function of time.  Due to experimental drift, the microwave frequency was $6$~kHz off resonance for these data as well as the data in Fig.~3A. Taking the mean of the left well data, and the peak value of the right well data, we conclude that there is a $0.84(2)$ probability that the atoms are in different spin states, and hence distinguishable, at the peak of the Rabi oscillations. The measured Rabi frequency in the right well is $32.05(18)~\rm kHz$, which is consistent with the measured oscillation frequency $32.6(6)\rm~kHz$ of the data in Fig.~3A.

\subsection{Summary of measurements and statistical analysis}
\label{sec:numberstable}

In the main text we present five distinct measurements, each of which contains a statistically-significant measurement of the HOM effect.  Table~\ref{tab:experimental_parameters} summarizes these experimental measurements and their uncertainties.
The quantities and their uncertainties are determined as follows:  $P_{\rm loss}$ is given by an average over all single-particle data in the relevant measurement; the uncertainty is the standard error determined from this set of points.  $A_{P_{11}}$, $A_{P_{\rm dist}}$,  $A_1$, $A_2$, $\phi$, and $P_{\rm HOM}^{\rm min}$ (Figs.~3A,B) are determined from weighted fits to the data, where each data point is weighted based on the statistical uncertainty associated with the measured value ($w_i = \frac{1}{\sigma_i^2}$), and the uncertainty in the final fit parameters are then calculated from the weighted average variance of the data.  $U/J$ and its uncertainty is determined from the procedure described in Sec.~\ref{sec:interaction}.  $P_{\rm HOM}^{\rm min}$ in Fig.~3C is determined from the data point at $\delta_{\rm cool} = 0$, and the associated uncertainty indicated by the error bar on this point, namely the standard error from the average over the experimental realizations (of which there are $\approx360$).

The purple circles shown in the figures in the manuscript represent $P_{\rm dist}$ as calculated from the expression $P_{\rm L}^1P_{\rm R}^2+P_{\rm R}^1P_{\rm L}^2$.  The error bars on these points are determined by propagating the uncertainty from each single-particle data point, properly taking into account correlations between these measurements.  Note that near the crossing point of the single-particle datasets the uncertainty in the calculated $P_{\rm dist}$ values is relatively small due to the fact that $P_{\rm dist}$ is the sum of two nearly anti-correlated variables and because the quantity reaches a minimum at this point.

In our analysis of the results presented in Figs.~2E,2F we must determine from our measurements of the amplitude of the $P_{\rm dist}(t)$ and $P_{11}(t)$ oscillations our confidence that these amplitudes are statistically different.  To do this we employ a modified Student's t-\-test known as Welch's t-\-test that is used for samples with possibly unequal variances and degrees of freedom. We obtain the effective degrees of freedom and perform a two-tailed test on the distribution to obtain the probability that the two parameters are different. Using this procedure and the standard errors from the fits (Table~\ref{tab:experimental_parameters}), we find for the $J/2\pi=262$ Hz experiment ($J/2\pi=348$ Hz experiment) a $6.1\sigma$ ($5.5\sigma$) deviation between the $P_{\rm dist}(t)$ data and $P_{11}(t)$ oscillation amplitudes.  Note that these are quite close to what one would calculate from a simple analysis of the standard errors quoted in Table~\ref{tab:experimental_parameters}, but this method accounts for the finite number of data points.

\section{Theoretical analysis}
\label{theory}

\subsection{Analysis of the three-dimensional optical tweezer potential}
\label{sec:DVR}

The double-well potential formed by two optical tweezers in the experiment is a novel realization for tunneling experiments, possessing properties different from typical lattice experiments. The potential resulting from the experimental configuration described in Sec.~\ref{sec:tweezer} is
\begin{align}
\label{eq:GaussPot}V\left(\mathbf{r}\right)&=-\frac{V_0}{1+\frac{z^2}{z_0^2}}\exp\!\!\left(\!\!\frac{-2x^2}{w_0^2\left(1+\frac{z^2}{z_0^2}\right)}\!\!\right)\!\!\!\left[\!\exp\!\!\left(\!\!-\frac{2\left(y-a/2\right)^2}{w_0^2\left(1+\frac{z^2}{z_0^2}\right)}\right)\!\!+\!\exp\!\!\left(\!\!-\frac{2\left(y+a/2\right)^2}{w_0^2\left(1+\frac{z^2}{z_0^2}\right)}\right)\!\!\right]\,
\end{align}
where $V_0$ is the single-well potential depth, $w_0$ is the beam
waist, $z_0$ is the $z$-length scale, $a$ is the gaussian function
separation, and we have assumed zero bias between wells.  For atoms trapped in a cubic optical lattice, the potential is separable in Cartesian coordinates and hence calculation of single-particle wave functions is straightforward.  In contrast, the optical tweezer potential of Eq.~(\ref{eq:GaussPot}) is not separable, and a quantitative treatment of the single-particle dynamics requires a numerical solution of the full 3D Schr\"odinger equation.

We solve for the eigenstates of Eq.~\eqref{eq:GaussPot} numerically using a 3D discrete variable representation (DVR)~\cite{DVR}.  As Eq.~\eqref{eq:GaussPot} is invariant under the inversion of any Cartesian coordinate, the eigenstates can be classified according to their parities $\left(P_x,P_y,P_z\right)$, where, e.g., $P_x=+$(-) denotes an eigenstate that is even (odd) with respect to inversion along the $x$ direction.  The DVR basis has been adapted to take into account this parity symmetry.  Additionally, due to the multiplicative separability of Eq.~\eqref{eq:GaussPot} along $x$ and $y$, we use a quasi-adiabatic method to reduce the effort required to converge the lowest-energy states.  That is, we first diagonalize the $y$ kinetic energy together with Eq.~\eqref{eq:GaussPot} at $x=0$ and a range of $z$ specified by the $z$-coordinate DVR basis.  For each $z$ we then project the $y$ kinetic energy and the potential into a truncated basis of the lowest-energy states of the 1D problem and diagonalize the $x$ and $y$ kinetic energies together with Eq.~\eqref{eq:GaussPot}.  After projecting the Hamiltonian for the $x$ and $y$ degrees of freedom into a truncated basis of the lowest-energy states of the 2D problem, we diagonalize the full 3D problem in this reduced basis.  Computational effort is reduced for fixed error using the quasi-adiabatic method because the quasi-adiabatic basis sizes are typically much smaller than the DVR basis sizes. Our results have been checked for convergence in both the DVR basis size and the basis size of the quasi-adiabatic decomposition.  

Once we have the eigenstates with parity $(P_x,P_y,P_z)$, $\psi_{P_x,P_y,P_z;n}\left(\mathbf{r}\right)$, and their eigenenergies
$E_{P_x,P_y,P_z;n}$, where $n$ labels eigenstates increasing in energy as $n=1,2,\dots$, we choose the basis states of our effective model, Eq.(~\ref{eq:Heff}) below, as
\begin{align}
\label{eq:L}\psi_R\left(\mathbf{r}\right)&=\frac{1}{\sqrt{2}}\left(\psi_{+++;1}\left(\mathbf{r}\right)+\psi_{+-+;1}\left(\mathbf{r}\right)\right)\\
\label{eq:R}\psi_L\left(\mathbf{r}\right)&=\frac{1}{\sqrt{2}}\left(\psi_{+++;1}\left(\mathbf{r}\right)-\psi_{+-+;1}\left(\mathbf{r}\right)\right)\, .
\end{align}
The choice of wave functions in Eqs.~\eqref{eq:L}-\eqref{eq:R} follows from truncating to the two lowest-energy states, which is valid as long as interactions and temperature are smaller than the energy separation from the excited states. We then choose the particular linear combinations of these states that are maximally localized around the left or right well.  Localized basis functions minimize the effect of interactions between wells in the resulting model.  For the parameters of the present experiment, interactions between particles in different wells are smaller than 1\% of the tunneling, and so are neglected.

From the calculated wave functions, we can determine the single-atom tunnel-coupling $J$ between the tweezers and
the two-atom interaction $U$.  The tunneling and intra-well interaction matrix elements are defined as
\begin{align}
J&=(E_{+-+;n}-E_{+++;n})/2\, ,\\
U&=\frac{4\pi \hbar^2 a_s}{m}\int d\mathbf{r} \left|\psi_{L}\left(\mathbf{r}\right)\right|^4\, ,
\end{align}
where  $a_s$=5.45 nm is the background $s$-wave scattering length for $^{87}$Rb~\cite{Rbas}, and $U$ is defined for two atoms in the left tweezer.

%%%%%%%%%%%%%%%%%%%%%%%%%%%%%%%%%%%%%

\subsection{Analysis of the two-particle dynamics}
\label{sec:tpd}

With the choice of single-particle basis states Eqs.~\eqref{eq:L}-\eqref{eq:R}, the effective Hamiltonian for two particles in the tunneling phase of the experiment is
\begin{align}
\label{eq:Heff}\hat{H}&=\left(\begin{array}{cccc} U&-2J&0&0\\-2J&0&0&0\\ 0&0&U&0\\ 0&0&0&0\end{array}\right)\, ,
\end{align}
where the two-particle basis is
\begin{align}
\langle \mathbf{r}_1,\mathbf{r}_2 |+\rangle&=\frac{1}{\sqrt{2}}\left(\psi_L\left(\mathbf{r}_1\right)\psi_L\left(\mathbf{r}_2\right)+\psi_R\left(\mathbf{r}_1\right)\psi_R\left(\mathbf{r}_2\right)\right)\\
\langle\mathbf{r}_1,\mathbf{r}_2|S\rangle&=\frac{1}{\sqrt{2}}\left(\psi_L\left(\mathbf{r}_1\right)\psi_R\left(\mathbf{r}_2\right)+\psi_R\left(\mathbf{r}_1\right)\psi_L\left(\mathbf{r}_2\right)\right)\\
\langle \mathbf{r}_1,\mathbf{r}_2|-\rangle&=\frac{1}{\sqrt{2}}\left(\psi_L\left(\mathbf{r}_1\right)\psi_L\left(\mathbf{r}_2\right)-\psi_R\left(\mathbf{r}_1\right)\psi_R\left(\mathbf{r}_2\right)\right)\\
\langle\mathbf{r}_1,\mathbf{r}_2|A\rangle&=\frac{1}{\sqrt{2}}\left(\psi_L\left(\mathbf{r}_1\right)\psi_R\left(\mathbf{r}_2\right)-\psi_R\left(\mathbf{r}_1\right)\psi_L\left(\mathbf{r}_2\right)\right)\, .
\end{align}

The time-dependent probability to measure one atom in each well,
\begin{align}
\label{eq:PLR}P_{11}\left(t\right)&=\mathrm{Tr}\left[\hat{\rho}\left(t\right) \left(|S\rangle\langle S|+|A\rangle\langle A|\right)\right]\, ,
\end{align}
where $\hat{\rho}\left(t\right)$ is the density matrix at time $t$ with initial condition
\begin{align}
\hat{\rho}\left(0\right)&=\left(\begin{array}{cccc}
a&\alpha&\beta&\gamma\\
\alpha^{\star}&b&\epsilon&\zeta\\
\beta^{\star}&\epsilon^{\star}&c&\eta\\
\gamma^{\star}&\zeta^{\star}&\eta^{\star}&1-a-b-c\end{array}\right)\,,
\end{align}
is
\begin{align}
\label{eq:PLRt}P_{11}\left(t\right)&=1-a-c+\frac{4J\left(2J\left(a-b\right)+U\mathcal{R}\left(\alpha\right)\right)}{\omega_{JU}^2}-\frac{4J\left(2J\left(a-b\right)+U\mathcal{R}\left(\alpha\right)\right)}{\omega_{JU}^2}\cos\left(\omega_{JU}t\right)\\
&-\frac{4 J \mathcal{I}\left(\alpha\right)}{\omega_{JU}}\sin\left(\omega_{JU}t\right)\, .
\end{align}
Here $\omega_{JU}^2=16J^2+U^2$, and $\mathcal{R}\left(\bullet\right)$ and
$\mathcal{I}\left(\bullet\right)$ denote real and imaginary parts.
This expression will be used in the next section to understand the
dynamics of two atoms in the presence of interactions.
%%%%%%%%%%%%%%%%%%%%%%%%%%%%%%%%%%%%%

\subsection{Calculation of $P_{11}$ and contrast for {\it distinguishable} atoms}

\label{sec:bounds}
% Here we derive constraints on the dynamics of $P_{11}(t)$, bounding both its minimum value and the amplitude of its oscillations, under the assumption that the two atoms are distinguishable.  The bound, referred to as $P_{\rm dist}$, is violated by the experimental data, and thus the following discussion justifies our claim of indistinguishability.  In the analysis below we write simplified expressions for the situation of zero particle loss.  In all descriptions in the main text and in the final plots in Fig.~\ref{fig:Finite_U_Bound} the loss is included.
Here we derive constraints on the dynamics of $P_{\rm dist}(t)$---the theoretically expected value of $P_{11}(t)$ for distinguishable particles---bounding both its minimum value and the amplitude of its oscillations.  The experimentally measured amplitude ($A_{P_{11}}$) and HOM dip ($P_{\rm HOM}^{\rm min}$) violate these bounds, and thus the following discussion justifies our claim of indistinguishability.  In the analysis below we write simplified expressions for the situation of zero particle loss.  In all descriptions in the main text and in the final plots in Fig.~\ref{fig:Finite_U_Bound} the loss is included.

\subsubsection{Single-atom density matrices}

By assumption of distinguishability, we can label an atom initially in the left well as ``atom 1'' and an atom initially in the right well as ``atom 2'', regardless of whether there is one or two atoms.  We define the point in time immediately after the ramp down of the double-well barrier to be $t=0$, at which point the density matrices for atoms 1 and 2 are
\begin{equation}
\def\arraystretch{1.1}
\rho_{1}(0)=\overbrace{\left( \begin{array}{cc}
\rho^{\mathrm{LL}}_{1}(0) & \rho^{\mathrm{LR}}_{1}(0) \\
\rho^{\mathrm{RL}}_{1}(0) & \rho^{\mathrm{RR}}_{1}(0)\\ \end{array}
\right)}^{\mathrm{originally~in~left~well}},~~~~~~~~\def\arraystretch{1.1}
\rho_2(0)=\overbrace{\left( \begin{array}{cc}
\rho^{\mathrm{LL}}_{2}(0) & \rho^{\mathrm{LR}}_{2}(0) \\
\rho^{\mathrm{RL}}_{2}(0) & \rho^{\mathrm{RR}}_{2}(0)\\ \end{array} \right)}^{\mathrm{originally~in~right~well}}.\nonumber
\end{equation}

These density matrices can be visualized on a Bloch sphere (see Fig.~\ref{fig:Bloch_Sphere}) by associating the $z$ direction with the $|L\rangle,~|R\rangle$ basis, and choosing the direction defined by
the ground state of the single-atom tunneling Hamiltonian
[$(|{\rm L}\rangle+|{\rm R}\rangle)/\sqrt{2}$] to be the $x$ direction, such that
\begin{equation}
\def\arraystretch{1.1}
\rho_{1}(0)=\frac{1}{2}\left( \begin{array}{cc}
1+z_1& x_1-iy_1 \\
x_1+iy_1 & 1-z_1\\ \end{array}
\right),~~~~~~~~\def\arraystretch{1.1}
\rho_2(0)=\frac{1}{2}\left( \begin{array}{cc}
1+z_2& x_2-iy_2 \\
x_2+iy_2 & 1-z_2\\ \end{array}
\right).\nonumber
\end{equation}
In this picture, the single particle dynamics amounts to rotation about the $x$-axis.  The single-atom experiments measure the probability $P_{\rm L}=\rho^{\rm LL}(t)$ for an atom starting in either the left or right well to
be in the left well at time $t$.  The oscillation amplitude of $P_{\rm L}$ for an atom starting in the left (right) well is $A_1$
($A_2$), and this amplitude determines the length of the Bloch vector's projection into the $yz$ plane
\begin{equation}
\label{eq:amplitudes}
A_{1}=\sqrt{y_1^2+z_1^2}~~~~~~A_{2}=\sqrt{y_2^2+z_2^2}.
\end{equation}
The experimentally measured phases $\phi_1$ and $\phi_2$, which can be nonzero due to small but non-vanishing tunneling initiated during the final stages of the ramp (right before $t=0$) determine the angle of the Bloch vectors in the
$yz$ plane at $t=0$, measured from
the $z$ axis,
\begin{equation}
\label{eq:phi}
\tan\phi_1=y_1/z_1~~~~~~\tan\phi_2=y_2/z_2.
\end{equation}
Equations \eqref{eq:amplitudes} and \eqref{eq:phi}, taken together,
are sufficient to reconstruct $y_{1(2)}$ and $z_{1(2)}$ from the
experimental data.  The data do {\it not}, however, specify the initial $x$-component of
the Bloch vector, and this unknown degree of freedom must be
considered in the presence of interactions (this issue is considered
in detail below).  In the experiment, the single-atom
observables are the two diagonal elements $\rho_1^{\mathrm{LL}}(t)$ and
$\rho_2^{\mathrm{LL}}(t)$, reported as the blue and red traces
in Figs.~2B,C of the manuscript, respectively.  In terms of the
contrast and phase, we have
\begin{eqnarray}
\rho_{1}^{\rm LL}(t)&=&\frac{1}{2}+\frac{A_{1}}{2}\cos(2Jt-\phi_1),\\
\rho_{2}^{\rm LL}(t)&=&\frac{1}{2}-\frac{A_{2}}{2}\cos(2Jt-\phi_2).
\end{eqnarray}

\subsubsection{Inferred two-atom dynamics in the absence of interactions}

In the experiment, the atoms are prepared independently in spatially separated tweezers, and they do not interact appreciably before $t=0$.  As a result, they must be uncorrelated at the onset of the tunneling, and we therefore consider the dynamics of an initial product state of distinguishable atoms
\begin{equation}
\label{eq:tpsdm}
\rho(0)=\rho_1(0)\otimes\rho_2(0).
\end{equation}
If interactions are ignored during the tunneling, it can be shown that
\begin{eqnarray}
P_{\rm dist}(t)&\equiv&\mathrm{Tr}\left[\rho(t)|\mathrm{L}\rangle_1\langle
    L|_1\otimes|\mathrm{R}\rangle_2\langle
    \mathrm{R}|_2\right]
+\mathrm{Tr}\left[\rho(t)|\mathrm{R}\rangle_1\langle
    R|_1\otimes|\mathrm{L}\rangle_2\langle
    \mathrm{L}|_2\right]\nonumber\\
&=&\rho_1^{\rm LL}(t) \rho_2^{\rm RR}(t)+\rho_1^{\rm RR}(t) \rho_2^{\rm LL}(t)\nonumber\\
&=&P^{\rm L}_1(t) P^{\rm R}_2(t)+P_1^{\rm R}(t) P_2^{\rm L}(t)
\label{P11dist}
\end{eqnarray}
satisfies
\begin{equation}
P_{\rm dist}(t)=\frac{1}{2}+\frac{A_1A_2}{2}\cos(2J t-\phi_1)\cos(2J t-\phi_2).
\label{suppMFFp11}
\end{equation}
Equation (\ref{suppMFFp11}) constrains that distinguishable and uncorrelated particles exhibit oscillations in $P_{11}$ with a contrast $A_1A_2/2$. 

 As discussed in Sec.~\ref{sec:preparation}, nonzero phase shifts of $\phi_{1,2}=2 J
t_{\rm corr}$ are measured in the experiment.  Thus the intersection of $\rho_{1}^{\mathrm{LL}}(t)$ and
$\rho_2^{\mathrm{LL}}(t)$ occurs at a time
$t_{\mathrm{HOM}}=2\pi/8J+t_{\rm corr}$, and
$\rho_{1}^{\mathrm{LL}}(t_{\mathrm{HOM}})=\rho_{2}^{\mathrm{LL}}(t_{\mathrm{HOM}})=1/2$ (neglecting single-atom loss).  At this time, we therefore have
\begin{equation}
\cos(2J t_{\mathrm{HOM}} -\phi_1)=\cos(2J
t_{\mathrm{HOM}}-\phi_2)=0,
\end{equation}
and hence $P_{\rm dist}(t_{\mathrm{HOM}})=1/2$.  This result is not surprising, because two uncorrelated and distinguishable atoms, which are each equally likely to be found in either well, ought to be found in different wells with probability $1/2$.  Thus, for noninteracting particles, an experimental measurement of $P_{11}(t_{\mathrm{HOM}})<1/2$ is a direct demonstration of indistinguishability.

\subsubsection{Inferred two-atom dynamics with interactions\label{sec:interactions}}

The two-particle dynamics in the presence of interactions depends not only on the measured Bloch vector components $y_{1(2)}$ and $z_{1(2)}$, but also on the unknown values $x_{1(2)}$.  Though undetermined, these coherences are constrained to satisfy
\begin{equation}
|x_1|\leq(1-A_{1}^2)^{1/2}~~~{\rm and}~~~|x_2|\leq(1-A_{2}^2)^{1/2}.
\end{equation}
These equations simply enforce that the total single-particle Bloch vector length must be less than unity, and $x_{1(2)}$ take on their extremal values only for a pure state.

The dynamics for finite $U$, parameterized by the unknown coherences $x_1$ and $x_2$, can be calculated for the initial two-particle density matrix in Eq.~(\ref{eq:tpsdm}) by using the results of Sec.~\ref{sec:tpd}.  After some algebra, we find that the minimum value of $P_{\rm dist}(t)$ is given by
\begin{align}
\mathcal{P}(x)=&\frac{2+A_1A_2}{4}+\frac{U(Jx+UA_1A_2\cos(2\phi)/4)}{16J^2+U^2}\nonumber\\
-&\frac{J\sqrt{(Ux-4JA_1A_2\cos2\phi)^2+64J^2A_1^2A_2^2\cos^2\phi\sin^2\phi+U^2A_1^2A_2^2\sin^2(2\phi)}}{16J^2+U^2},
\label{p11}
\end{align}
while the contrast is given by
\begin{align}
\mathcal{A}(x)=&\frac{2J\sqrt{(U x-4JA_1A_2\cos2\phi)^2+64J^2A_1^2A_2^2\cos^2\phi\sin^2\phi+U^2A_1^2A_2^2\sin^2(2\phi)}}{16J^2+U^2}.
\label{Ap11}
\end{align}
Here $x=x_1+x_2$, and we have set $\phi_1=\phi_2=\phi$ (these
phases are measured to be very nearly equal in the experiment).  In
principle, the unknown coherences $x_1$ and $x_2$ can be nonzero.  If
so, these would necessarily develop at the end of the ramp of the trap to the tunneling parameters [time
interval $T_2$ in Fig.~\ref{fig:tunneling_prep}A] as a result of this ramp speed not being
infinitely fast compared to the tunneling $J$.  However, we note that
the development of coherences due to quasi-adiabatic effects during
the ramp down requires a bias $\Delta\neq0$.  For a fixed bias, a
simple model of the Landau-Zener type crossing predicts that
$x_1=-x_2$, and thus $x=0$.  If, on the other hand, the bias is
fluctuating, then we also expect these coherences to remain zero.
Therefore, it is exceedingly likely that $x=0$ at the onset of the experiment, resulting in the bounds
\begin{equation}
\label{eq:bound_natural}
P_{\rm dist}(t_{\rm HOM})\geq \mathcal{P}(0)~~~~{\rm and}~~~~A_{P_{\rm dist}}\leq\mathcal{A}(0).
\end{equation}

A more conservative bound can be derived by ignoring our reasonable expectation that $x=0$ at the onset of tunneling, and simply minimizing [maximizing] $\mathcal{P}(x) [\mathcal{A}(x)]$ over $x$.  This occurs when $x$ takes on the extremal value
\begin{equation}
x_0=-(1-A_1^2)^{1/2}-(1-A_2^2)^{1/2}
\end{equation}
(i.e. when both atoms start in a pure state in a direction diametrically opposed to the ground state of
the tunneling Hamiltonian), thus providing more conservative bounds on the distinguishable expectation
\begin{equation}
\label{bound}
P_{\rm dist}(t_{\rm HOM})\geq \mathcal{P}(x_0)~~~~{\rm and}~~~~A_{P_{\rm dist}}\leq\mathcal{A}(x_0).
\end{equation}

These bounds on $P_{\rm dist}(t_{\rm HOM})$ and $A_{P_{\rm dist}}$ are plotted as black solid ($x=0$) and red dotted ($x=x_0$) lines in Fig.~\ref{fig:Finite_U_Bound}.  In Figs.~\ref{fig:Finite_U_Bound}A,B we see that, while interactions do allow $P_{\rm dist}(t_{\rm HOM})$ to fall below $1/2$, the experimentally measured HOM dip ($P_{\rm HOM}^{\rm min}$, blue point) still falls below the minimal value allowable for distinguishable particles.  In Figs.~\ref{fig:Finite_U_Bound}C,D we see that interactions can also increase the oscillation contrast of $P_{\rm dist}(t)$, but the experimentally measured contrast ($A_{P_{11}}$, blue point) still falls above the maximum value allowed for distinguishable particles.  We emphasize again that the minimization [maximization] of $\mathcal{P}(x) [\mathcal{A}(x)]$ over $x$ is extremely conservative, because the types of states that minimize it---pure states with coherences diametrically opposed to the direction set by the tunnel coupling---are highly implausible.  Focusing on the more natural bounds given in Eq.~(\ref{eq:bound_natural}), we see that interactions actually tend to {\it increase} the minimum possible value of the HOM dip for distinguishable particles, while {\it decreasing} the maximum contrast.

The experimentally measured single-particle quantities used for producing the plots in Fig.~\ref{fig:Finite_U_Bound}, and the uncertainties propagated through Eqs.~(\ref{p11},\ref{Ap11}) to determine the error bands, are shown in Table \ref{tab:experimental_parameters}.

\clearpage

\noindent {\bf \Large Figures and Tables}

\

\

\begin{figure}[h]
	\centering
	\includegraphics[scale=.9]{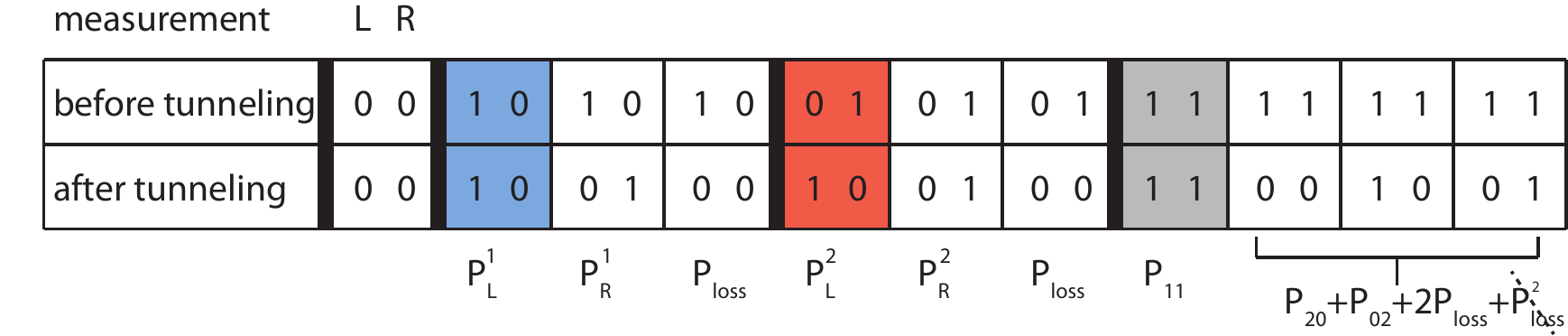}
	\caption{{\bf A list of the possible outcomes of our imaging protocol.}   A 0 (1) within a box indicates that on the pixel corresponding to either the left ($L$) or right ($R$) wells, the measured counts fell below (exceeded) the threshold for triggering atom detection.  The red, blue, and grey regions highlight the signals used to produce the data points in, for example, Fig.~2B,E. $P_i^{1}$($P_i^{2}$) refers to an atom that started on the left (right), i.e. the first image indicated an atom on the left (right).}
	\label{fig:imaging}
\end{figure} 

\begin{figure}[h!]
	\centering
	\includegraphics[scale=.53]{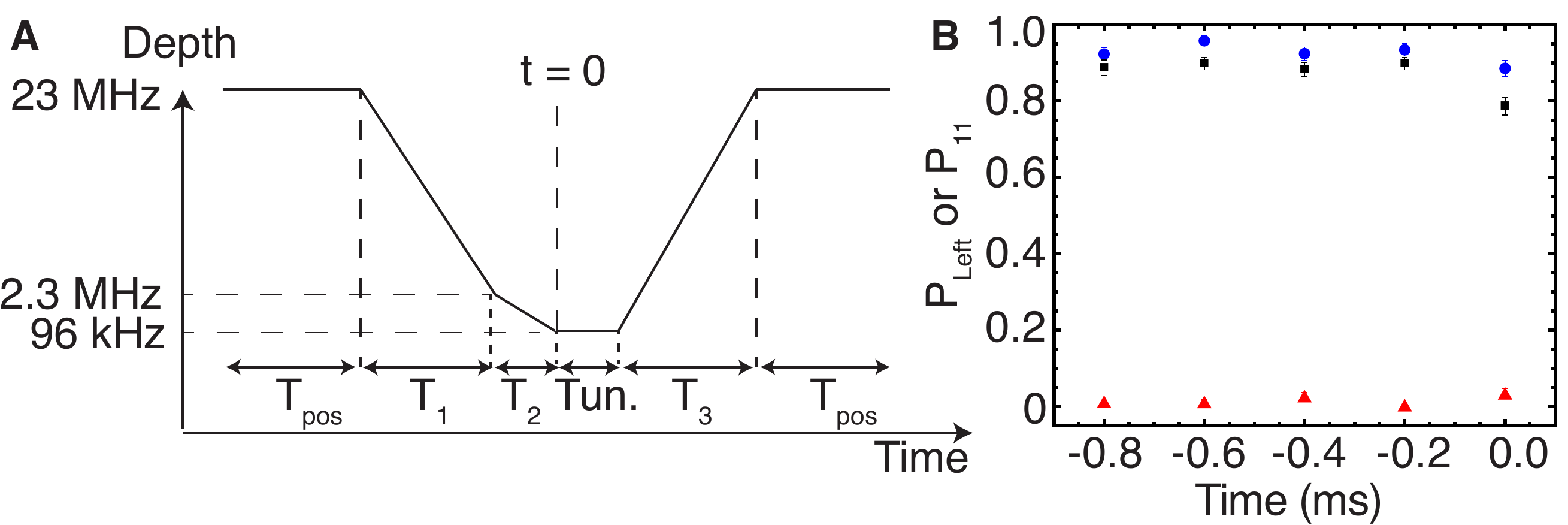}
	\caption{{\bf Protocol for initiating tunneling. (A)} The tunneling sequence as a function of time, illustrated for the 96 kHz final trap depth. {\bf (B)} For the 96 kHz depth, tunneling at times before $t=0$ for a single atom starting on the right (red), on the left (blue), and with one atom in each well (black).}
	\label{fig:tunneling_prep}
\end{figure} 

\begin{figure}[h!]
	\centering
	\includegraphics[scale=.31]{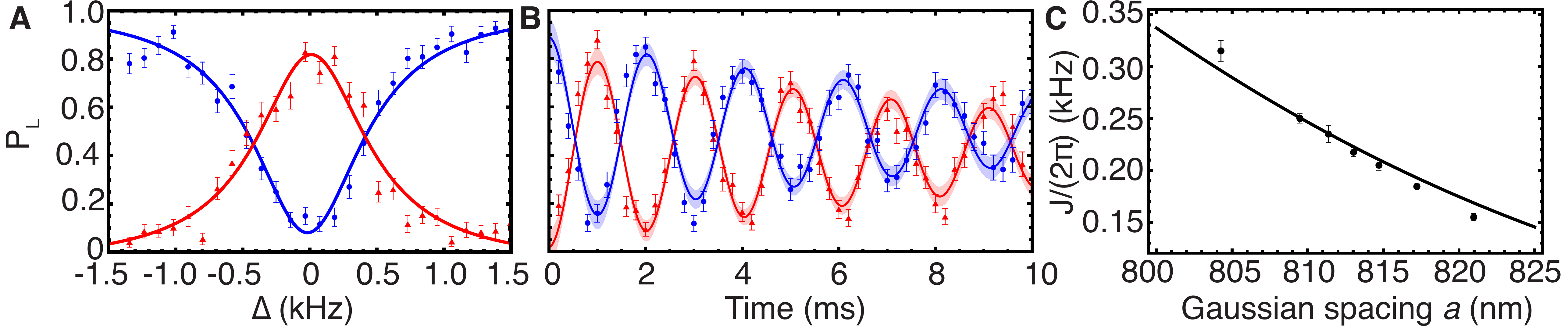}
	\caption{\textbf{Tunneling dependences at the 96 kHz well depth. (A)} For a 0.9 ms tunneling time, we scan the relative well bias, $\Delta$, and observe the single particle tunneling resonance, symmetrically for an atom originating from either well. The blue circles (red triangles) correspond to atoms starting in the left (right) well. \textbf{(B)} At $\Delta=0$, we observe oscillations at $2J$ in the expectation value of an atom's position. \textbf{(C)} We measure and plot the tunneling $J$ as a function of the gaussian function spacing $a$. The solid black line is a theoretical expectation for a 96 kHz single well depth and 707 nm gaussian spot size.}
	\label{fig:tunneling}
\end{figure} 

\begin{figure}[h!]
	\centering
	\includegraphics[scale=.45]{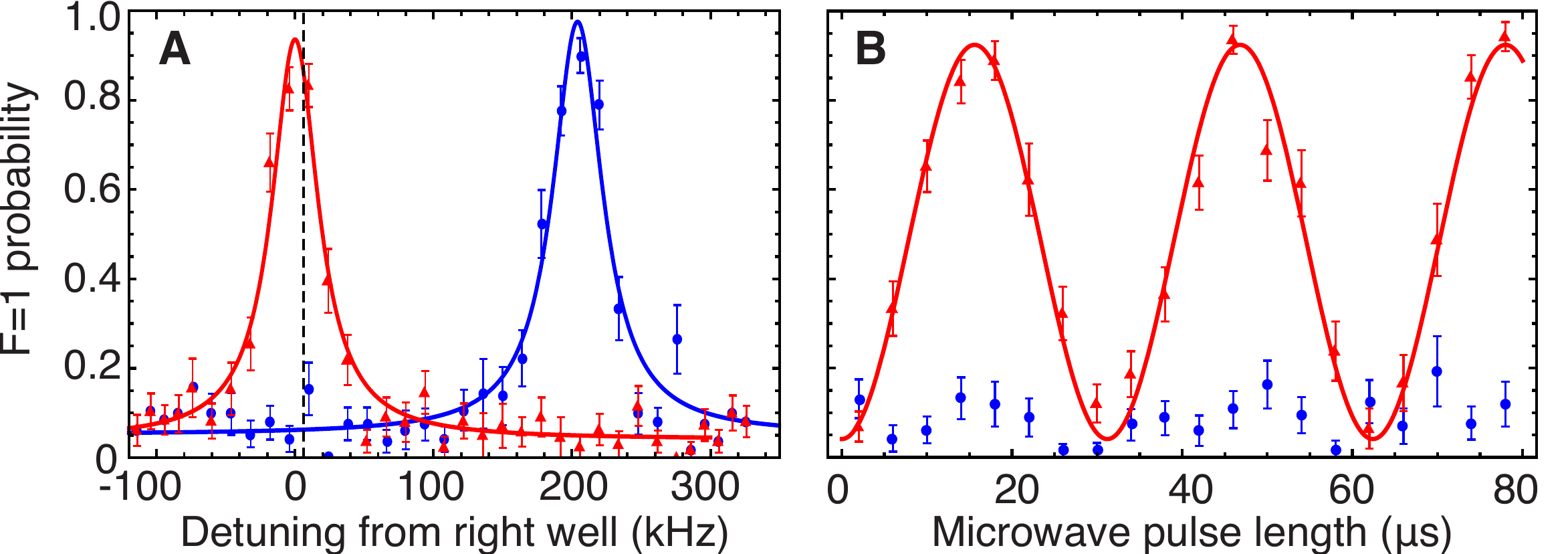}
	\caption{\textbf{Single atom spin-flip data.}  After the protocol we use to flip a single spin, we measure the spin state with a resonant push-out beam that removes $F=2$ atoms from the trap~\cite{Kaufman2012Supp}. \textbf{(A)} Microwave spectrum with a $16~\mu$s microwave pulse for the right well (red triangles) and left well (blue circles). The dashed line indicates the detuning used in the experiment in Fig.~3A.  \textbf{(B)} Microwave Rabi oscillations for a detuning of $+6$kHz from the right well resonance (indicated by the dashed line in A) for the right well (red triangles) and left well (blue circles).}
	\label{fig:spinFlip}
\end{figure} 

\begin{figure}[h!]
\centering
\includegraphics[width=12cm]{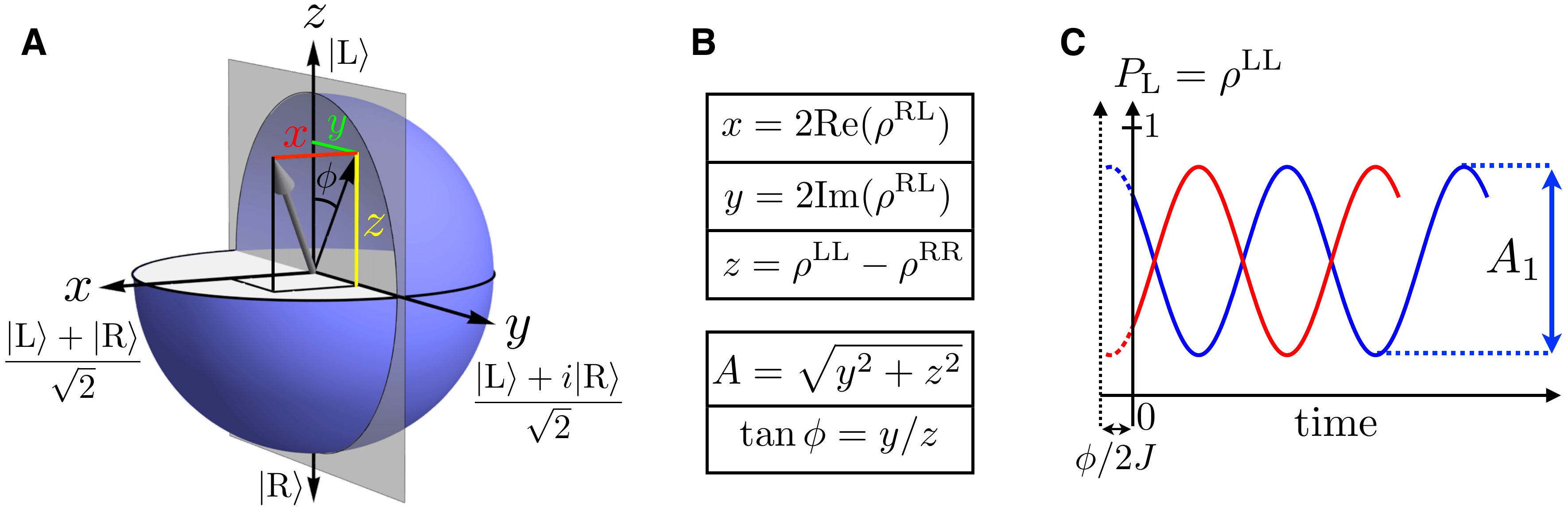}
\caption{\textbf{(A)} Bloch sphere used to describe the single-atom density
  matrices. \textbf{(B)} Two tables: The top one summarizes the connection between Cartesian
  coordinates of the Bloch vector and populations/coherences in the
  $|{\rm L}\rangle$,$|{\rm R}\rangle$ basis, while the bottom one connects the
  Cartesian coordinates to experimentally measured phase and
  contrast. \textbf{(C)} Schematic of the single-particle dynamics and the
  meaning of $\phi_{1(2)}$ and $A_{1(2)}$.}
\label{fig:Bloch_Sphere}
\end{figure}

\begin{figure}[h!]
\centering
\includegraphics[width=12 cm]{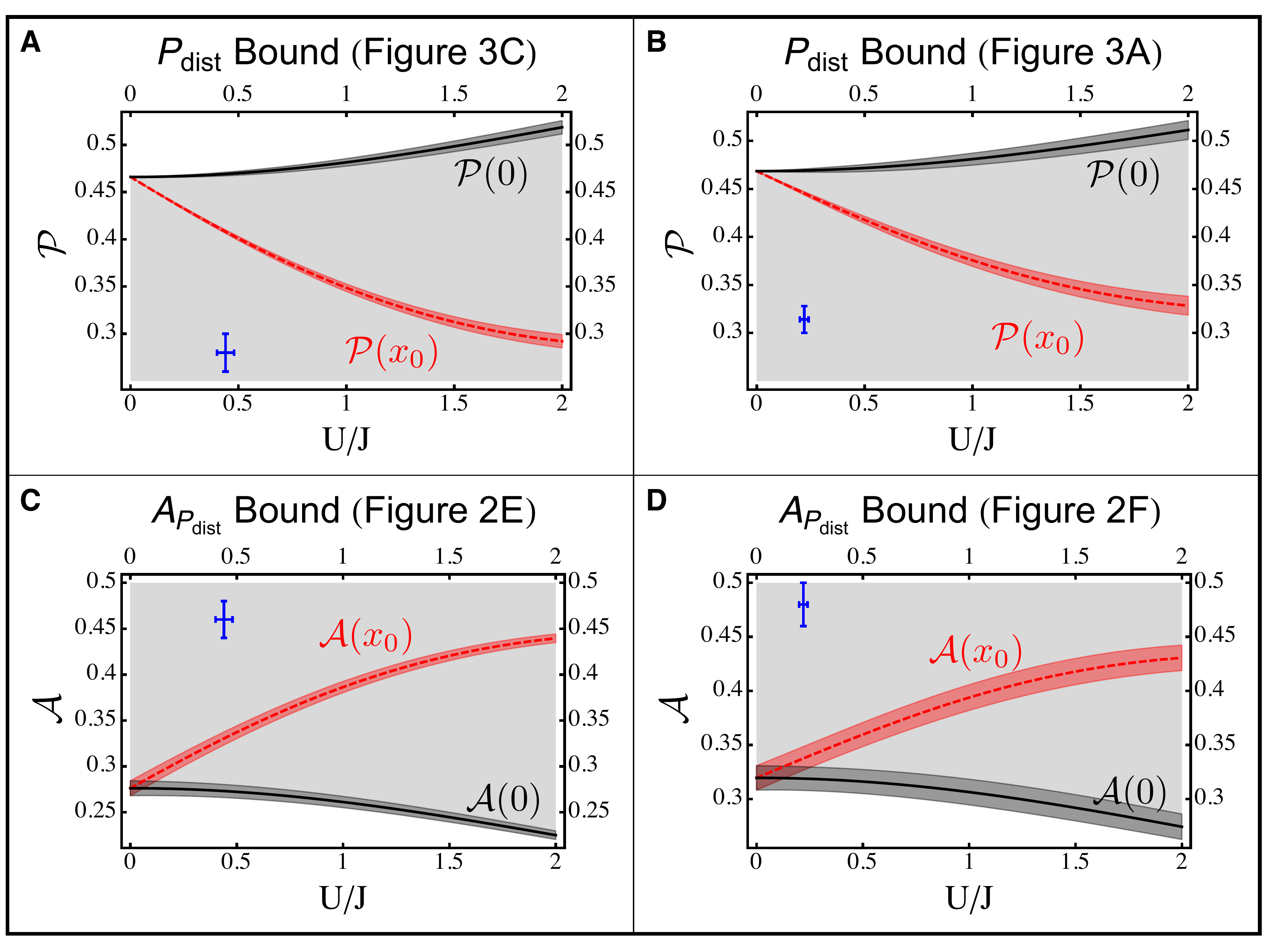}
\caption{\textbf{Bounds on $P_{\rm dist}(t_{\rm HOM})$ (A,B) and $A_{P_{\rm dist}}$ (C,D)}:  The two left panels (A,C) are for experiments with $U/J=0.44(4)$, while the two right panels (B,D) are for experiments with $U/J=0.22(2)$.  In all plots, the figure of the manuscript to which the presented data corresponds is given in the plot label.  In panels A and B, the black curve is $\mathcal{P}(x=0)$, which is a lower bound on the HOM dip for distinguishable particles assuming no initial coherences along the $x$-direction of the Bloch sphere.  The dark shaded uncertainty region of the curve is obtained by propagating uncertainties in the experimentally measured single-particle  amplitudes ($A_1$,$A_2$) and phases ($\phi_1$,$\phi_2$) through Eq.~\eqref{p11}.  The light shaded region below the black curve is therefore classically forbidden, i.e. inaccessible to distinguishable particles.  The red dashed curve is $\mathcal{P}(x_0)$, which is a worst case scenario that could, in principal, be saturated by distinguishable particles (the red shaded region is obtained in the same way).  The measured minimum of the HOM dip ($P_{\rm HOM}^{\rm min}$, blue point) sits in the classically forbidden region. Panels C and D are similar to the top panels, except now we plot the contrast bound ($\mathcal{A}$) and the blue points are measured values of $A_{P_{11}}$.}
\label{fig:Finite_U_Bound}
\end{figure}

\clearpage

\begin{table}[hp]
		\centering
	\begin{tabular}{l c c c c c c c}
	 & & & & & & & \\
	%\hline
	\multicolumn{8}{c}{Important numbers from tunneling dynamics plots} \\
	%\cline{2-6}
	\hline
	\multicolumn{1}{| l ||}{Measurement} & $P_{\text{loss}}$ & $A_{P_{11}}$ & $A_{P_\text{dist}}$  & $U/J$ & $A_{1}$ & $A_{2}$ & \multicolumn{1}{c |}{$\phi$} \\ 
	\hline %\hline
	\multicolumn{1}{| l ||}{Fig.~2B, 2E} & $0.049(2) $ & $0.46(2)$ & $0.282(12)$ & $0.44(4)$ & $0.722(15)$ & $0.765(16)$ & \multicolumn{1}{c |}{$-0.20(7)$} \\ 
	\multicolumn{1}{| l ||}{Fig.~2C, 2F} & $0.034(2) $ & $0.48(2)$ & $0.306(18)$ & $0.22(2)$ & $0.77(2)$ & $0.83(2)$ & \multicolumn{1}{c |}{$-0.45(7)$} \\ 
	\hline
	 & & & & & & & \\
	%\multicolumn{2}{c}{Item} \\
	%\cline{1-2}
	\multicolumn{4}{c}{Important numbers from $t_{\rm HOM}$ plots} \\
	\cline{1-4}
	 \multicolumn{1}{| c ||}{Measurement} &  \multicolumn{1}{ c |}{ $P_{\text{loss}}$} &  \multicolumn{1}{ c |}{$(1-P_{\rm loss})^2/2$}  & \multicolumn{1}{ c |}{$P^{\rm min}_{HOM}$ }& & &  \\
	\cline{1-4}	
	 \multicolumn{1}{| l ||}{Fig.~3A} &\multicolumn{1}{ c |}{ $0.032(2) $ }&\multicolumn{1}{ c |}{ $0.4689(19)$} & \multicolumn{1}{ c |}{$0.314(14)$} & & & & \\ 
	 \multicolumn{1}{| l ||}{Fig.~3B} &\multicolumn{1}{ c |}{ $0.0337(17) $ }&\multicolumn{1}{ c |}{ $0.4669(16)$} & \multicolumn{1}{ c |}{$0.296(10)$} & & & & \\ 
	 \multicolumn{1}{| l ||}{Fig.~3C} &\multicolumn{1}{ c |}{ $0.0346(15) $ }&\multicolumn{1}{ c |}{ $0.4660(14)$} & \multicolumn{1}{ c |}{$0.28(2)$} & & & & \\ 
	\cline{1-4}
	\end{tabular}
		\caption{\label{tab:experimental_parameters} 
		Summary of the important parameters for understanding the experimentally observed two-particle interference.   The experimental signature of the HOM effect is reflected in the difference between our measurements and the expectation for distinguishable atoms:  In the top table the values to compare are $A_{P_{11}}$  and $A_{P_\text{dist}}$.  In the bottom table the values to compare are $(1-P_{\rm loss})^2/2$ and $P^{\rm min}_{\rm HOM}$. Each value of $P_{\rm loss}$ is computed by taking the mean over all single atom data in the specified set. The values tabulated here are used to produce the distinguishable limits shown in Fig.~\ref{fig:Finite_U_Bound}. 
		}
	\end{table}

\

\

\bibliographystyle{Science}

\begin{thebibliography}{10}

\bibitem{Glauber1963}
R.~J. Glauber, {\it Phys. Rev.\/} {\bf 130}, 2529 (1963).

\bibitem{Twiss}
{R. Hanbury Brown and R. Q. Twiss}, {\it Nature\/} {\bf 177}, 27 (1956).

\bibitem{HOM}
C.~K. Hong, Z.~Y. Ou, L.~Mandel, {\it Phys. Rev. Lett.\/} {\bf 59}, 2044
  (1987).

\bibitem{Beugnon2006}
J.~Beugnon, {\it et~al.\/}, {\it Nature\/} {\bf 440}, 779 (2006).

\bibitem{Walraff2013}
C.~Lang, {\it et~al.\/}, {\it Nat. Phys.\/} {\bf 9}, 345 (2013).

\bibitem{electron}
E.~Bocquillon, {\it et~al.\/}, {\it Science\/} {\bf 339}, 1054 (2013).

\bibitem{Vuletic1998}
V.~Vuletic, C.~Chin, A.~J. Kerman, S.~Chu, {\it Phys. Rev. Lett.\/} {\bf 81},
  5768 (1998).

\bibitem{Weiss2004}
D.~S. Weiss, {\it et~al.\/}, {\it Phys. Rev. A\/} {\bf 70}, 040302(R) (2004).

\bibitem{MeschedePNAS}
A.~Steffen, {\it et~al.\/}, {\it Proceedings of the National Academy of
  Sciences\/} {\bf 109}, 9770 (2012).

\bibitem{Yasuda1996}
M.~Yasuda, F.~Shimizu, {\it Phys. Rev. Lett.\/} {\bf 77}, 3090 (1996).

\bibitem{Folling2005}
S.~F\"{o}lling, {\it et~al.\/}, {\it Nature\/} {\bf 434}, 481 (2005).

\bibitem{Schellekens2005}
M.~Schellekens, {\it et~al.\/}, {\it Science\/} {\bf 310}, 648 (2005).

\bibitem{Ottl2005}
A.~Ottl, S.~Ritter, M.~Kohl, T.~Esslinger, {\it Phys. Rev. Lett.\/} {\bf 95},
  090404 (2005).

\bibitem{Rom2006}
T.~Rom, {\it et~al.\/}, {\it Nature\/} {\bf 444}, 733 (2006).

\bibitem{Jeltes2007}
T.~Jeltes, {\it et~al.\/}, {\it Nature\/} {\bf 445}, 402 (2007).

\bibitem{Hodgman2011}
S.~S. Hodgman, R.~G. Dall, A.~G. Manning, K.~G.~H. Baldwin, A.~G. Truscott,
  {\it Science\/} {\bf 331}, 1046 (2011).

\bibitem{MilburnRM}
P.~Kok, {\it et~al.\/}, {\it Rev. Mod. Phys.\/} {\bf 79}, 135 (2007).

\bibitem{Giovannetti1}
V.~Giovannetti, S.~Lloyd, L.~Maccone, {\it Science\/} {\bf 306}, 1330 (2004).

\bibitem{Vetsch2010}
E.~Vetsch, {\it et~al.\/}, {\it Phys. Rev. Lett.\/} {\bf 104}, 203603 (2010).

\bibitem{Thompson2013}
J.~D. Thompson, {\it et~al.\/}, {\it Science\/} {\bf 340}, 1202 (2013).

\bibitem{Schlosser2001}
N.~Schlosser, G.~Reymond, I.~Protsenko, P.~Grangier, {\it Nature\/} {\bf 411},
  1024 (2001).

\bibitem{Jochim2011}
F.~Serwane, {\it et~al.\/}, {\it Science\/} {\bf 332}, 336 (2011).

\bibitem{Kaufman2012}
A.~M. Kaufman, B.~J. Lester, C.~A. Regal, {\it Phys. Rev. X\/} {\bf 2}, 041014
  (2012).

\bibitem{Monroe1995}
C.~Monroe, {\it et~al.\/}, {\it Phys. Rev. Lett.\/} {\bf 75}, 4011 (1995).

\bibitem{Supplement}
See supplementary materials.

\bibitem{Folling2007}
S.~F\"{o}lling, {\it et~al.\/}, {\it Nature\/} {\bf 448}, 1029 (2007).

\bibitem{Anderlini2007}
M.~Anderlini, {\it et~al.\/}, {\it Nature\/} {\bf 448}, 452 (2007).

\bibitem{Strabley}
J.~Sebby-Strabley, {\it et~al.\/}, {\it Phys. Rev. Lett.\/} {\bf 98}, 200405
  (2007).

\bibitem{Weitenberg2011}
C.~Weitenberg, {\it et~al.\/}, {\it Nature\/} {\bf 471}, 319 (2011).

\bibitem{Bakr2009}
W.~S. Bakr, J.~I. Gillen, A.~Peng, S.~F\"{o}lling, M.~Greiner, {\it Nature\/}
  {\bf 462}, 74 (2009).

\bibitem{Ma2011}
R.~Ma, {\it et~al.\/}, {\it Phys. Rev. Lett.\/} {\bf 107}, 095301 (2011).

\bibitem{Egorov2013}
M.~Egorov, {\it et~al.\/}, {\it Phys. Rev. A\/} {\bf 87}, 053614 (2013).

\end{thebibliography}

\begin{thebibliography}{10}

\bibitem{RauschenbeutelLightShift}
F.~Le~Kien, P.~Schneeweiss, A.~Rauschenbeutel, {\it EPJ D\/} {\bf
  67}, 92 (2013).

\bibitem{Kaufman2012Supp}
A.~M. Kaufman, B.~J. Lester, C.~A. Regal, {\it Phys. Rev. X\/} {\bf 2}, 041014
  (2012).

\bibitem{Schlosser2001Supp}
N.~Schlosser, G.~Reymond, I.~Protsenko, P.~Grangier, {\it Nature\/} {\bf 411},
  1024 (2001).

\bibitem{Bakr2009Supp}
W.~S. Bakr, J.~I. Gillen, A.~Peng, S.~F\"{o}lling, M.~Greiner, {\it Nature\/}
  {\bf 462}, 74 (2009).

\bibitem{Weitenberg2011Supp}
C.~Weitenberg, {\it et~al.\/}, {\it Nature\/} {\bf 471}, 319 (2011).

\bibitem{Sompet2013}
P.~Sompet {\it et~al.},
\newblock {\it Phys. Rev. A} {\bf 88}, 051401(R) (2013).

\bibitem{DVR}
D.~T.~Colbert and W.~H.~Miller, {\it J. Chem. Phys.} {\bf 96}, 1982 (1992).

\bibitem{Rbas}
P. S. Julienne, F. H. Mies, E. Tiesinga, and C. J. Williams, {\it Phys. Rev. Lett.} {\bf 78}, 1880 (1997).




\end{thebibliography}

\end{widetext}

\end{document}